\documentclass[12pt,english]{article}
\usepackage[english]{babel}
\usepackage{amsmath,amssymb,amscd,xcolor}
\usepackage{amsfonts}
\usepackage{mathrsfs}
\usepackage{mathtools}
\usepackage{float}
\usepackage{graphicx}
\usepackage{url}
\usepackage{hyperref}
\usepackage{cite}
\usepackage{physics}
\usepackage{braket}
\usepackage{verbatim}
\usepackage[font=footnotesize,labelfont=bf,width=.9\textwidth]{caption}
\usepackage{multirow}
\usepackage{boldline}
\usepackage{enumitem}
\usepackage{wrapfig}

\usepackage{tikz}
\usetikzlibrary{decorations.markings, shapes.misc}
\usetikzlibrary{arrows}
\tikzset{->-/.style={decoration={markings, mark=between positions 0.0 and 1.0 step 1cm with {\arrow{stealth}}},postaction={decorate}}}

\usepackage{scalerel}
\usepackage[normalem]{ulem}
\allowdisplaybreaks

\hypersetup{
    colorlinks,%
    citecolor=blue,%
    filecolor=blue,%
    linkcolor=blue,%
    urlcolor=blue
}

\makeatletter
\renewcommand\section{\@startsection {section}{1}{\z@}%
                                   {-5.5ex \@plus -1ex \@minus -.2ex}
                                   {2.3ex \@plus.2ex}%
                                   {\normalfont\large\bfseries}}
\renewcommand\subsection{\@startsection{subsection}{2}{\z@}%
                                     {-3.25ex\@plus -1ex \@minus -.2ex}%
                                     {1.5ex \@plus .2ex}%
                                     {\normalfont\bfseries}}

\numberwithin{equation}{section}

\newcommand{\subalign}[1]{%
  \vcenter{%
    \Let@ \restore@math@cr \default@tag
    \baselineskip\fontdimen10 \scriptfont\tw@
    \advance\baselineskip\fontdimen12 \scriptfont\tw@
    \lineskip\thr@@\fontdimen8 \scriptfont\thr@@
    \lineskiplimit\lineskip
    \ialign{\hfil$\m@th\scriptstyle##$&$\m@th\scriptstyle{}##$\hfil\crcr
      #1\crcr
    }%
  }%
}
\newcommand{\vast}{\bBigg@{4}}
\newcommand{\Vast}{\bBigg@{5}}
\makeatother

\newcommand{\bea}{\begin{eqnarray}}
\newcommand{\eea}{\end{eqnarray}}
\newcommand{\beq}{\begin{equation}}
\newcommand{\eeq}{\end{equation}}
\newcommand{\be}{\begin{equation}}
\newcommand{\ee}{\end{equation}}

\newcommand{\mf}[1]{\mathfrak{#1}}
\newcommand{\ms}[1]{\mathsf{#1}}

\newcommand{\mc}[1]{\mathcal{#1}}

\newcommand{\mTr}{\mathrm{Tr}}

\newcommand{\Li}[1]{\text{Li}_{#1}}
\newcommand{\gev}[1]{\big\langle #1\big\rangle_{\rm grav}}
\newcommand{\barmc}[1]{\bar{\mathcal{#1}}}

\newcommand{\Zs}{Z_\text{scalar}}
\newcommand{\Zg}{\mathcal{Z}_\text{grav}}

\newcommand{\mss}{\mathsf{s}}
\newcommand{\su}{\mathfrak{su}(2)}
\newcommand{\slalg}{\mathfrak{sl}(2, \mathbb{R})}
\newcommand{\so}{\mathfrak{so}(1,3)}
\newcommand{\hypgeo}[4]{{}_2F_1\left( #1, #2; #3; #4 \right)}
\newcommand{\lds}{\ell_{\text{dS}}}
\newcommand{\lads}{\ell_{\text{AdS}}}

\newcommand{\SCS}{S_{\text{CS}}}
\newcommand{\msh}{\mathsf{h}}
\newcommand{\msR}{\mathsf{R}}

\renewcommand{\title}[1]{\vbox{\center\LARGE{#1}}\vspace{5mm}}
\renewcommand{\author}[1]{\vbox{\center#1}\vspace{5mm}}
\newcommand{\address}[1]{\vbox{\center\footnotesize\em#1}}
\newcommand{\email}[1]{\vbox{\center\footnotesize\tt#1}\vspace{5mm}}

\begin{document}
\begin{titlepage}
 \begin{flushright}

\end{flushright}

\begin{center}

\hfill \\
\hfill \\
\vskip 1cm
\title{Spinning up the spool:\\
Massive spinning fields in 3d quantum gravity}

\author{Robert Bourne, Alejandra Castro, and Jackson R. Fliss}
\address{
{Department of Applied Mathematics and Theoretical Physics, University of Cambridge,\\ Cambridge CB3 0WA, United Kingdom}
}
  
\email{rjb260@cam.ac.uk, ac2553@cam.ac.uk, jf768@cam.ac.uk}

\end{center}

\vfill
\abstract{

We show how to incorporate massive spinning fields into the Euclidean path integral of three-dimensional quantum gravity via its Chern-Simons formulation. The coupling of the spinning fields to gravity is captured by a Wilson spool, a collection of Wilson loops winding around closed paths of the geometry, and generalizes the proposal of \cite{Castro:2023dxp,Castro:2023PRL}.  We present a robust derivation of the Wilson spool by providing a new group-theoretic perspective of the quasinormal mode method for one-loop determinants. We test our proposal on Euclidean BTZ and $S^3$ backgrounds. We also evaluate explicitly the quantum corrections to the path integral on $S^3$, and report on how $G_N$ and the mass are renormalized to leading order in perturbation theory. 

}
\vfill

\end{titlepage}

\newpage
{
  \hypersetup{linkcolor=black}
  \tableofcontents
}

\section{Introduction}

Low-dimensional gravity is an exciting arena to explore and test the gravitational path integral. In two and three spacetime dimensions, there is no propagating graviton and all of the effective degrees of freedom are long-range. A prime example of this phenomenon is the rewriting of pure Einstein gravity with a cosmological constant (of either sign) as a Chern-Simons gauge theory \cite{Achucarro:1987vz, Witten:1988hc} which is the quintessential example of a topological field theory in three-dimensions. A full leveraging of this fact allows the exact evaluation of the gravitational path integral either about a saddle point \cite{Banados:1998tb, Govindarajan:2002ry,Cotler:2018zff,Anninos:2020hfj,Collier:2023fwi} or as a non-perturbative sum over saddles \cite{Carlip:1992wg,Guadagnini:1994ahx,Witten:2007kt,Castro:2011xb}. While Chern-Simons gravity is not a UV-complete model of quantum gravity \cite{Witten:2007kt,Maloney:2007ud,Castro:2011xb}, its all-loop exactness provides strong tests for potential microscopic models in the spirit of e.g. \cite{Banerjee:2011jp,Sen:2012dw,Bhattacharyya:2012ye,Sen:2012kpz,Giombi:2013fka,Anninos:2020hfj}.

One feature that is expected of a UV-complete model of quantum gravity is that it includes matter, in particular massive fields that couple to gravity. The manifest topological invariance that makes Chern-Simons theory so powerful as a description of the gravitational path integral also presents a challenge to incorporating matter. On a practical level, this is simple to illustrate: the action of a massive field theory minimally coupled to a geometry involves both inverse metrics and metric determinants. The rewriting of these terms as Chern-Simons connections is highly non-linear and indicates that integrating out of the massive field will result in a non-local effective action. However, we can take inspiration from the general philosophy that the low-energy avatar of the worldline of a massive degree of freedom is a line-operator of the effective gauge theory \cite{Ooguri:1999bv}.

This philosophy was made precise in \cite{Castro:2023dxp} for massive scalar fields minimally coupled to gravity with a positive cosmological constant. The key result was the expression of the one-loop determinant of a massive scalar field coupled to a background metric, $g_{\mu\nu}$, as a gauge-invariant object of the Chern-Simons connections, $A_{L/R}$. That is,  
\beq\label{eq:spoolsketch}
    Z_\text{scalar}[g_{\mu\nu}]=\exp\frac{1}{4}\mathbb W[A_L,A_R]~.
\eeq
The object $\mathbb W[A_L,A_R]$, coined the Wilson spool, is a collection of Wilson loop operators wrapped many times around cycles of the base geometry. The equality in \eqref{eq:spoolsketch} is expected to apply to three-dimensional gravity of either sign of cosmological constant, and this was explicitly shown for Euclidean black holes in Anti-de Sitter (i.e., Euclidean BTZ) and Euclidean de Sitter (i.e., the three-sphere $S^3$) in \cite{Castro:2023dxp,Castro:2023PRL}. It has also been upheld on $T\bar T$ deformations of AdS$_3$ \cite{He:2024pbp}. The importance of \eqref{eq:spoolsketch} is not only conceptual, it is practical: it was additionally shown in \cite{Castro:2023dxp} that certain ``exact methods" in Chern-Simons theory (such as Abelianisation \cite{Blau:1993tv,Blau:2006gh,Blau:2013oha} and supersymmetric localization \cite{Kapustin:2009kz,Marino:2011nm}) extend to three-dimensional de Sitter (dS$_3$) Chern-Simons gravity with the Wilson spool inserted into the path integral. This allows a precise and efficient calculation of the quantum gravitational corrections to $Z_\text{scalar}$ at any order of perturbation theory of Newton's constant, $G_N$.

An important element excluded from the dictionary established in \cite{Castro:2023dxp} is the description of massive fields with spin. The representation theory of dS$_3$ spacetime admits a series of representations describing single-particle states with both mass and integer spin \cite{Joung:2006gj,Sun:2021thf}. It is important to know how the field theory of those particles fits into quantum gravity about the dS$_3$ background. This consideration is more than pedestrian: we expect the massive states of a UV description of quantum gravity to carry spin and it is sensible to organize these states into the representation theory of the low energy effective theory. In \cite{Castro:2023PRL} a Wilson spool describing massive spinning fields on Euclidean BTZ was conjectured and shown to reproduce the correct one-loop determinants as $G_N\rightarrow0$, however, no principled derivation was provided there. More importantly, it was shown that a na\"ive generalization of the Wilson spool does not correctly reproduce the physics of spinning fields on $S^3$.

In this paper we address this mismatch by revisiting the derivation of $\mathbb W[A_L,A_R]$. The original derivation in \cite{Castro:2023dxp} to obtain \eqref{eq:spoolsketch} used some of the conditions behind the quasinormal mode method to evaluate one-loop determinants developed in \cite{Denef:2009kn}. Here, we give a robust derivation where we cast all aspects of the method in terms of conditions that rely solely on group-theoretic concepts.  These conditions apply to gravity with either sign of cosmological constant, $\Lambda$. We show in explicit examples how from the group-theoretic picture one can write the one-loop determinant in terms of $(A_L,A_R)$ and representations of the algebra. This then gives the Wilson spool, which describes the minimal coupling of massive spinning fields to gravity.

To be specific, we state our main result, the generalization of \eqref{eq:spoolsketch} for massive spinning fields,  as the following. Consider the local path integral,\footnote{Including any additional St\"uckelberg fields to fix its invariances and associated ghosts.} $Z_{\Delta,\mss}$, of a spin-$\mss$ field $\Phi_{\mu_1\mu_2\ldots\mu_s}$ with mass
\beq
    \frac{m^2}{\Lambda}=(\Delta+\mss-2)(\mss-\Delta)~,
\eeq
minimally coupled to a metric geometry, $(M_3,g_{M_3})$,  where $M_3$ is topologically either Euclidean BTZ or Euclidean dS$_3$. Then, we propose 
\beq\label{eq:preMR}
    \log Z_{\Delta,\mss}[g_{M_3}]=\frac{1}{4}\mathbb W_{j_L,j_R}[A_L,A_R]~,
\eeq
where
\begin{multline}\label{eq:MainResult}
    \mathbb W_{j_L,j_R}[A_L,A_R]=\frac{i}{2}\int_{\mc C}\frac{\dd\alpha}{\alpha}\frac{\cos\left(\frac\alpha2\right)}{\sin\left(\frac\alpha2\right)}\left(1+\mss^2\sin^2\left(\frac\alpha2\right)\right)\times\\
    \sum_{\msR_L\otimes\msR_R}\mTr_{\msR_L}\left(\mc Pe^{\frac{\alpha}{2\pi}\oint_\gamma A_L}\right)\mTr_{\msR_R}\left(\mc Pe^{-\frac{\alpha}{2\pi}\oint_\gamma A_R}\right)~.
\end{multline}
The details of $\mathbb W_{j_L,j_R}$ will be made explicit as we continue, however let us briefly summarize the parts appearing in \eqref{eq:MainResult}. The Chern-Simons connections, $A_{L/R}$, are related to the metric, $g_{M_3}$, in \eqref{eq:preMR} through the usual Chern-Simons gravity dictionary and they are integrated over a non-trivial cycle, $\gamma$, of the base geometry. The representations, $\msR_{L/R}$, appearing in the Wilson loops are summed over a set determined by the mass and spin, $(\Delta,\mss)$, of \eqref{eq:preMR} and labelled by weights $(j_L,j_R)$. Lastly the parameter $\alpha$ is integrated along a contour, $\mc C$, determined by a regularization scheme appropriate for the sign of cosmological constant. The ultimate effect of the $\alpha$ integral is to implement a ``winding" of the Wilson loop operators around $\gamma$; this occurs through the summing the residues of the poles of its measure (as well as any of representation traces themselves). We coin the above object, \eqref{eq:MainResult}, the {\it spinning Wilson spool.}

In the remainder of this paper we will establish the results upholding \eqref{eq:preMR} and \eqref{eq:MainResult} as well as the necessary details to utilize \eqref{eq:MainResult}. Much of this analysis will be done in the context of a positive cosmological constant, where the difference between the scalar and spinning Wilson spools are most apparent. In Sec.\,\ref{sec:nonStandardReps} we will establish the representation theory necessary for describing massive spin-$\mss$ particles in terms of the Euclidean isometry group $\su_L\oplus\su_R$. In Sec.\,\ref{section:SpoolDerivation} we will utilize this representation theory to build the spinning Wilson spool, \eqref{eq:MainResult}, when the background is classically $S^3$. This follows from a re-organization of the quasinormal mode method for one-loop determinants \cite{Denef:2009kn} in terms of representation theory following a set of broad guiding principles, which we establish in Sec.\,\ref{subsect:su2QMNs} and apply to Chern-Simons gravity in Sec.\,\ref{subsect:spoolcons}. In Sec.\,\ref{sec:QGcorr} we propose the spinning Wilson spool as an operator, allowing us to insert it off-shell into the Chern-Simons path integral. The full power of this is manifest in the context of a dS$_3$ background where we show that the Chern-Simons path integral with a spool insertion reduces to an ordinary integral that can be evaluated to any order in $G_N$ perturbation theory. We illustrate how this can be interpreted as either a renormalization of the particle mass or Newton's constant, depending on choice of renormalization condition. Lastly in Sec.\,\ref{sec:AdSspool}, to establish the scope of the results in Sec.\,\ref{section:SpoolDerivation}, we will also illustrate how \eqref{eq:MainResult} also holds when the background is Euclidean BTZ {\it mutatis mutandis}. We conclude the paper in Sec.\,\ref{sec:disc} with potential uses for the spinning Wilson spool as well as a discussion of open problems. Details on the dictionary relating Chern-Simons theory and gravity as well as details on the spin-$\mss$ quasinormal modes on $S^3$ are found in App.\,\ref{app:conv} and App.\,\ref{app:DHSspinS3}, respectively.

\section{Non-standard spinning representations of \texorpdfstring{$\mf{su}(2)$}{su(2)}}\label{sec:nonStandardReps}

In this section, we will describe single-particle states of massive spin-$\mss$ fields living on  dS$_3$, with de Sitter radius $\lds$, as representations of $\su_L\oplus\su_R$. 
The guiding principle of the construction is to mimic the unitary representations of the Lorentzian dS$_3$ isometry group, $\so$. 
In \cite{Castro:2023dxp} this was done for cases with $\mss=0$, and here, following the same arguments and conventions, we extend that construction to include spin.

Unitary representations of $\so$ are labeled by a conformal dimension $\Delta$ (the eigenvalue of the dilatation operator $D$) and a spin $\mss$ (the eigenvalue of the spin operator $iM$) and come in two series:\footnote{See \cite{Castro:2023dxp} for a basic introduction on $\so$ representation theory and for our notation here. See \cite{Joung:2006gj, Joung:2007je, Basile:2016aen, Sun:2021thf, Penedones:2023uqc} for a more thorough treatment.}
\begin{align}\label{eq:so31Delta-s}
    \Delta=&1+\nu~,&&\nu\in(-1,1)~,&&\mss=0~,\nonumber\\
    \Delta=&1-i\mu~,&&\mu\in\mathbb{R}~,&&\mss\in\mathbb Z~.
\end{align}
The first line is the {\it complementary series} describing light scalars with masses $\lds^2m^2=1-\nu^2$ while the second line is the {\it spinning principal series} describing spinning particles with masses $\lds^2m^2=(\mss-1)^2+\mu^2$.

To connect this to $\su_L\oplus\su_R$, let us introduce some basic aspects of $\su$. The algebra is generated by $L_3$ and $L_\pm$, where  
\beq
[L_3,L_\pm]=\pm L_\pm~,\qquad [L_+,L_-]=2L_3~. 
\eeq
The Casimir of the representation is $c_2^{\su}=\frac{1}{2}(L_+L_-+L_-L_+)+L_3^2$. Representations of $\su$ will be characterized by  $j$, the eigenvalue of $L_3$.
The $\so$ algebra shares a complexification with $\su_L\oplus\su_R$, 
such that the dilatation  and spin  generators can be identified with the Cartan elements of $\su$ as
\beq\label{eq:SOCartantoSUCartan}
    D=-L_3-\bar L_3~,\qquad\qquad M=iL_3-i\bar L_3~,
\eeq
respectively. Similar relations follow for the remaining generators, which can be found in \cite{Castro:2023dxp}. In \eqref{eq:SOCartantoSUCartan} we have distinguished the generators of $\su_R$ from those of $\su_L$ by an overbar.  Within this complexification,  the quadratic Casimir of $\so$ is equal to the sum of the $\su$ Casimirs:
\beq\label{eq:SOcastoSUcas}
    c_2^{\so}=-2c_2^{\su_L}-2c_2^{\su_R}~.
\eeq
Equation \eqref{eq:SOCartantoSUCartan} suggests that we need to look for $\su_L\oplus\su_R$  representations with
\beq\label{eq:DeltaStoJLJR}
    \Delta=-j_R-j_L~,\qquad\qquad \mss=j_R-j_L~,
\eeq
where $j_L$ is the eigenvalue of $L_3$ in $\su_L$, and similarly for $j_R$. 

The relations \eqref{eq:DeltaStoJLJR}, together with \eqref{eq:so31Delta-s}, indicate to us that we will need to construct representations with continuous and complex eigenvalues $(j_L,j_R)$. These do not fall into the standard finite-dimensional representation theory of $\su$. However in \cite{Castro:2020smu, Castro:2023dxp} it was shown how to construct an alternative inner product in $\su$ (or equivalently, an alternative notion of Hermitian conjugation) allowing for highest-weight representations with a continuous and complex weight while preserving norm-positivity. Such representations were coined {\it non-standard representations} in the latter reference. These representations are built from a highest weight state, $|j,0\rangle$, satisfying
\beq
    L_3|j,0\rangle=j|j,0\rangle~,\qquad\qquad L_+|j,0\rangle=0~,
\eeq
and $j\in \mathbb{C}$ is the weight of the representation. Acting with lowering operators defines 
\beq
    |j,p\rangle=N_{j,p}\left(L_-\right)^p|j,0\rangle~,
\eeq
for some normalizations $N_{j,p}$ that we will determine shortly. 

The key ingredient distinguishing the non-standard representations is a map, $\mc S$, between highest weight representations
\beq\label{eq:shadowmapdef}
    \mc S|j,p\rangle=|j^\ast,p\rangle~,
\eeq
where $j^\ast$ is the complex conjugate of $j$. This map is utilized in Hermitian conjugation in the following way:
\beq\label{eq:HermConjdef}
    L_3^\dagger:=\mc S^{-1}L_3\mc S~,\qquad L_\pm^\dagger:=-\mc S^{-1}L_\mp\mc S~.
\eeq
Note that $j\overset{\mc S}{\rightarrow}j^\ast$ is consistent with the action of $L_3^\dagger$.
In \cite{Castro:2023dxp}, norm-positive inner products were constructed for non-standard representations with $j=-\frac{1}{2}(1+\nu)$ or $j=-\frac{1}{2}(1-i\mu)$ (and $\mu,\nu\in\mathbb R$) which relate to the scalar complementary and principal series with $\mss=0$. Here we will relax this condition and show that a norm-positive inner product can be constructed for {\it any} complex $j$. 

We will first determine the normalizations $N_{j,p}$. We note that
\beq\label{eq:norm-1}
    \langle j,p-1|L_+|j,p\rangle=\frac{N_{j,p}}{N_{j,p-1}}p(2j+1-p)\langle j,p-1|j,p-1\rangle~,
\eeq
which can be seen from replacing $L_+L_-$ with $c_2^{\su}+L_3-L_3^2$ in the matrix element. Alternatively, utilizing \eqref{eq:HermConjdef} and \eqref{eq:shadowmapdef}, this same matrix element is
\beq\label{eq:norm-2}
    \langle j,p-1|L_+|j,p\rangle=\langle j,p|L_+^\dagger|j,p-1\rangle^\ast=-\left(\frac{N_{j^\ast,p-1}}{N_{j^\ast,p}}\right)^\ast \langle j,p|j,p\rangle~.
\eeq
If we assume that we can set $\langle j,p-1|j,p-1\rangle=1$ and also $\langle j,p|j,p\rangle=1$, then \eqref{eq:norm-1} and \eqref{eq:norm-2} imply 
\beq
    \left(\frac{N_{j^\ast,p-1}}{N_{j^\ast,p}}\right)^\ast=p(p-2j-1)\left(\frac{N_{j,p}}{N_{j,p-1}}\right)~.
\eeq
To solve this constraint, we write
\beq
    \alpha_{j,p}=\text{Arg}\left(N_{j,p}\right)~,\qquad \phi_{j,p}=\text{Arg}\left(p-2j-1\right)~,
\eeq
leading to recurrence relations
\begin{align}
    \frac{\abs{N_{j^\ast,p-1}}}{\abs{N_{j^\ast,p}}}=&p\abs{p-2j-1}\frac{\abs{N_{j,p}}}{\abs{N_{j,p-1}}}~,\nonumber\\
    \alpha_{j^\ast,p}-\alpha_{j^\ast,p-1}=&\alpha_{j,p}-\alpha_{j,p-1}+\phi_{j,p}\;\;\;\text{mod}(2\pi)~.
\end{align}
There is a lot of freedom in solving these recurrence relations. We will choose a particular solution by fixing 
\beq
    \abs{N_{j,p}}=\abs{N_{j^\ast,p}}~,\qquad\qquad \alpha_{j,0}=\alpha_{j^\ast,0}=0~,\qquad\qquad \alpha_{j,p}=-\alpha_{j^\ast,p}\;\;\forall\;p~.
\eeq
This sets
\beq
    \frac{\abs{N_{j,p-1}}}{\abs{N_{j,p}}}=\frac{\abs{N_{j^\ast,p-1}}}{\abs{N_{j^\ast,p}}}=\sqrt{p\abs{p-2j-1}}
\eeq
and\footnote{Note that $\phi_{j^\ast,p}=-\phi_{j,p}$.}
\beq
    \alpha_{j,p}=-\frac{1}{2}\sum_{p'=1}^p\phi_{j,p'}~.
\eeq
For concreteness we summarize the generator actions on this representation as
\begin{align}\label{eq:GeneratorActionsOnStates}
    L_3|j,p\rangle&=(j-p)|j,p\rangle\nonumber~,\\    L_-|j,p\rangle&=e^{i\phi_{j,p}/2}\sqrt{(p+1)\abs{p-2j}}|j,p+1\rangle~,\nonumber\\
    L_+|j,p\rangle&=-e^{i\phi_{j,p}/2}\sqrt{p\abs{p-2j-1}}|j,p-1\rangle~,
\end{align}
with the action on $|j^\ast,p\rangle$ obtained by simply replacing $j\rightarrow j^\ast$ in the above formulas. We can now show norm-positivity by induction starting with $\langle j,0|j,0\rangle=1$. Let us investigate the first descendant state:
\begin{align}
    \abs{L_-|j,0\rangle}^2=&-\langle j,0|\mc S^{-1}L_+\mc S L_-|j,0\rangle\nonumber\\
    =& e^{i\phi_{j,1}/2+i\phi_{j^\ast,1}/2}\sqrt{\abs{2j}}\sqrt{\abs{2j}}\langle j,0|j,0\rangle\nonumber\\
    =&\abs{2j}>0~.
\end{align}
Similarly norm-positivity of $|j,p\rangle$ induces norm-positivity of $|j,p+1\rangle$\footnote{We are assuming $j$ is a generic complex number and ignoring the potential for possible null states when $j\in\frac{1}{2}\mathbb N_0$. Of course, in these cases the representation simply terminates and we recover the standard, finite-dimensional, representations of $\su$.}:
\begin{align}
    \abs{L_-|j,p\rangle}^2=&-\langle j,p|\mc S^{-1}L_+\mc S L_-|j,p\rangle\nonumber\\
    =&e^{i\phi_{j,p+1}/2+i\phi_{j^\ast,p+1}/2}\sqrt{(p+1)\abs{p-2j}}\sqrt{(p+1)\abs{p-2j}}\langle j,p|j,p\rangle\nonumber\\
    =&(p+1)\abs{p-2j}\langle j,p|j,p\rangle>0~.
\end{align}
This establishes norm-positivity for all states in the representation. We additionally note that these representations have well-defined characters
\beq\label{eq:NSchar}
    \chi_j(z)=\mTr_j\left(e^{i2\pi z L_3}\right)=\sum_{p=0}^{\infty}e^{i2\pi z(j-p)}=\frac{e^{i\pi(2j+1)z}}{2i\sin(\pi z)}~.
\eeq

In \cite{Castro:2023dxp} the restrictions of the highest-weights, $j=-\frac{1}{2}(1+\nu)$ or $j=-\frac{1}{2}(1-i\mu)$, followed from Hermiticity of the $\su$ Casimirs, separately. In this paper we will take a more Lorentzian perspective, and impose reality of the $\so$ Casimir at the level of \eqref{eq:SOcastoSUcas}; doing so ties the $\su_{L/R}$ representations together and the highest weights take the generic form
\beq\label{eq:HWSpinPS}
    j_L=-\frac{1}{2}(1+\mss-i\mu)~,\qquad j_R=-\frac{1}{2}(1-\mss-i\mu)~.
\eeq 
Let us briefly show this now. Assuming that $j_L$ and $j_R$ are generically complex the condition for reality of the $\so$ Casimir is that
\beq\label{eq:SOCasreality1}
    \text{Im}\left(j_L(j_L+1)\right)=-\text{Im}\left(j_R(j_R+1)\right)~,
\eeq
or equivalently
\beq\label{eq:SOCasreality2}
    \text{Im}(j_L)\left(1+2\text{Re}(j_L)\right)=-\text{Im}(j_R)\left(1+2\text{Re}(j_R)\right)~.
\eeq
If both sides are simultaneously zero then we must have either $j\in\mathbb R$ or $j\in-\frac{1}{2}+i\mathbb R$ (for either $j_L$ or $j_R$) which lead to the complimentary and principal-type representations discussed in \cite{Castro:2023dxp}. More generally there will be a family of highest weight solutions satisfying \eqref{eq:SOCasreality2}. However, if we further insist\footnote{At this point we will impose this by hand; we will see later that in the quantum theory only representations with $j_L-j_R\in\mathbb Z$ contribute to the matter one-loop determinant.} that $j_L-j_R\in\mathbb R$ then it must be the case that
\beq
    \text{Im}(j_L)=\text{Im}(j_R)~,\qquad\qquad \text{Re}(j_L)+\text{Re}(j_R)=-1~,
\eeq
which lead to the highest-weights appropriate for the spinning principal series, \eqref{eq:HWSpinPS}. We will refer to such representations as the {\it spinning principal-type representations.} 

As noted already in \cite{Castro:2023dxp}, our non-standard characters can be massaged into the suggestive Lorentzian form via
\beq
    \chi_{j_L}(z_L)\chi_{j_R}(z_R)+\chi_{j^\ast_L}(z_L)\chi_{j^\ast_R}(z_R)=\frac{w^{\mss}q^{\Delta}+w^{-\mss}q^{\bar \Delta}}{(1-w^{-1}q)(1-wq)}~,
\eeq
where $q=e^{-i\pi (z_L+z_R)}$ and $w=e^{i\pi(z_L-z_R)}$ and we have identified \eqref{eq:DeltaStoJLJR} as well as $\bar\Delta=2-\Delta$. This matches the Harish-Chandra character, $\mTr\left(q^Dw^{iM}\right),$ for the $\so$ spinning principal series \cite{Basile:2016aen}. 

\section{Spinning spool on \texorpdfstring{$S^3$}{S3}}\label{section:SpoolDerivation}

In this section, we present the construction of the Wilson spool for massive spin-$\mss$ fields on $S^3$. That is, we will derive an expression for the one-loop determinant of these fields on $S^3$ in terms of the representations of $\su_L\otimes \su_R$ constructed in the previous section.   

To start, let us describe the path integral of a single massive spin-$\mss$ field, with no self-interactions. The local partition function for this theory contains a symmetric spin-$\mss$ tensor, $\Phi_{\mu_1\mu_2\ldots\mu_\mss}$, as well as a tower of St\"uckelberg fields which enforce that $\Phi_{\mu_1\mu_2\ldots\mu_\mss}$ is transverse and traceless \cite{Zinoviev:2001dt}:
\beq
\nabla^\nu\Phi_{\nu\mu_2\ldots\mu_\mss}={\Phi^\nu}_{\nu\mu_3\ldots\mu_\mss}=0~.
\eeq
As emphasized in \cite{Anninos:2020hfj}, on a compact manifold the path integral over symmetric, transverse, traceless (STT) tensors leads to non-local divergences which cannot be canceled by local counterterms. This path integral must be compensated by the path integral over the St\"uckelberg fields and ghosts which leave behind a finite product from integrating over normalizable zero modes. To that end, we write
\beq\label{eq:ZDs1}
Z_{\Delta,\mss}=Z_{\text{zero}}Z_{\text{STT}}~,
\eeq
where
\beq\label{eq:Spin_sPathIntegral}
Z_{\text{STT}}=\int[\mathcal D\Phi_{\mu_1\mu_2\ldots\mu_\mss}]_{\text{STT}}e^{-\frac{1}{2}\int\Phi \left(-\nabla^2_{(\mss)}+\lds^2\bar m^2_{\mss}\right)\Phi}~.
\eeq
Above $\nabla_{(\mss)}^2$ is the Laplace-Beltrami operator
\beq
\left[ \nabla_{(\mss)}^2 \Phi \right]_{\mu_1\mu_2\ldots\mu_\mss} = \nabla_\nu \nabla^\nu \Phi_{\mu_1\mu_2\ldots\mu_\mss}~,
\eeq
and $\bar m^2_\mss$ is an ``effective mass"
\beq 
    \lds^2\bar{m}_\mss^2 = \lds^2m^2 + 3\mss - \mss^2~,
\eeq
where we recall that $m^2$ is the standard mass parameter in dS$_3$ \cite{Anninos:2020hfj}, and related to the representation theory of Sec.\,\ref{sec:nonStandardReps} via
\beq 
    \lds^2m^2 = (\Delta + \mss - 2)(\mss-\Delta)~.
\eeq
The zero mode contribution in \eqref{eq:ZDs1} follows from counting conformal Killing tensor modes on $S^3$ and is given by \cite{Anninos:2020hfj}\footnote{This is true for $\mss\geq 2$ while for $\mss=0,1$ the product over $n$ is replaced by 1.}
\beq\label{eq:Zzero}
Z_\text{zero}=\left[(\Delta - 1)(\bar{\Delta} - 1)\right]^{\frac{\mss^2}{2}}\prod_{n=0}^{\mss-2} \left[(\Delta + n)(\bar{\Delta} + n)\right]^{-(n+1 + \mss)(n+1-\mss)}~,
\eeq
where we recall that $\bar\Delta = 2-\Delta$. 

In the following, we will show that
\beq\label{eq:logZisW}
\log Z_{\Delta,\mss}=\frac{1}{4}\mathbb W_{j_L,j_R}~,
\eeq
for fields on $S^3$.
That is, we will express $Z_{\Delta,\mss}$ as a function of the Chern-Simons connections which we will explicitly construct utilizing the non-standard representation theory of Sec.\,\ref{sec:nonStandardReps}.  While our construction takes place on a fixed classical background, we will see that $\mathbb W_{j_L,j_R}$ is an integral over gauge-invariant Wilson loop operators and naturally generalizes into an off-shell operator that can be inserted into the Chern-Simons path integral, which we will discuss in Sec\,\ref{sec:QGcorr}.

The roadmap to derive \eqref{eq:logZisW} is as follows. We will first focus on $Z_{\text{STT}}$. This determinant can be evaluated via the method of quasinormal modes pioneered by Denef, Hartnoll, and Sachdev (DHS) \cite{Denef:2009kn}. We will adapt this method such that each component has a group theoretic interpretation: we will show how defining properties of quasinormal modes can be translated to conditions on the representation of the fields. This follows \cite{Castro:2023dxp}, however, for spinning fields, we will take particular care with the role of global conditions (i.e., Euclidean solutions are regular and single-valued) in isolating physical contributions to the quasinormal mode product.
 The additional contribution of $Z_\text{zero}$, which is not part of the quasinormal mode product but crucial for maintaining locality of $Z_{\Delta,\mss}$, will permit a Schwinger parameterization of $\log Z_{\Delta,\mss}$, regularized by an $i\varepsilon$ prescription. This will organize the quasinormal mode sum into an integral over representation traces of the background holonomies. From this follows our main result.

\subsection{A group theory perspective on \texorpdfstring{$S^3$}{S3} quasinormal modes}\label{subsect:su2QMNs}

As the first step in our construction, we will recast the functional determinant
\beq\label{eq:Spin_sFunctionalDetDifferential}
    Z_\text{STT}=\det\left(-\nabla_{(\mss)}^2+\lds^2\bar m_\mss^2\right)^{-\frac12}
\eeq
in $\mf{su}(2)$ representation theoretic language. We recall that the DHS method instructs us to treat $Z_\text{STT}^2$ as meromorphic function of $\Delta$. Then, up to a holomorphic function, $Z_\text{STT}^2$ is equal to the product containing the same zeros and poles. Here $Z_\text{STT}$ only has poles on states satisfying $(-\nabla^2_{(\mss)}+\bar m_\mss^2)\Phi_{\mu_1\mu_2\ldots\mu_\mss}=0$. These are precisely the spin-$\mss$ quasinormal modes. In App.\,\ref{app:DHSspinS3}, we explicitly compute these modes and their product to obtain $Z_\text{STT}$ directly. Here we will reinterpret these modes in terms of $\mf{su}(2)$ representation theory to obtain an expression natural to the Chern-Simons theory formulation of gravity.

We note that the isometry algebra of the three-sphere is generated by two mutually commuting sets of $\su$ vector fields $\{\zeta_a\}$ and $\{\bar\zeta_b\}$ which are the infinitesimal left and right group actions acting on $S^3\simeq SU(2)$. On spin-$\mss$ STT tensors the Casimirs of their Lie derivatives, $\{\mc L_{\zeta_a}\}$ and $\{\mc L_{\bar\zeta_b}\}$, act as the Laplace-Beltrami operator \cite{Rubin:1984}: 
\beq 
    -2\delta^{ab}\left(
        \mathcal{L}_{\zeta_a}\mathcal{L}_{\zeta_b} + \mathcal{L}_{\bar{\zeta}_a}\mathcal{L}_{\bar{\zeta}_b}
    \right) \Phi_{\mu_1\mu_2\ldots\mu_\mss} = \left[ \nabla^2_{(\mss)} -\mss(\mss+1)\right] \Phi_{\mu_1\mu_2\ldots\mu_\mss}~.
\eeq
 Hence we can write \eqref{eq:Spin_sFunctionalDetDifferential} suggestively as
\beq\label{eq:Spin_sFunctionalDetGroup}
    Z_\text{STT} = \det\left(2c_2^{\su_L}+2c_2^{\su_R}  + \Delta(2-\Delta)-\mss^2\right)^{-\frac{1}{2}}~.
\eeq
Following the DHS methodology, we then expect $Z_\text{STT}^2$ to have pole contributions from states in $\su_L\oplus\su_R$ representations satisfying 
\beq\label{eq:DHS_CasimirRelation}
\left(-2c_2^{\su_L}-2c_2^{\su_R}\right)|\psi\rangle = \left[ \Delta(2-\Delta) -\mss^2 \right]|\psi\rangle~.
\eeq 
This is precisely the condition satisfied by the non-standard representations constructed in Sec.\,\ref{sec:nonStandardReps} with highest weights $(j_L, j_R) = (-\frac{\Delta+\mss}{2}, -\frac{\Delta-\mss}{2})$. We are interested in the poles in $Z_\text{STT}^2$ arising from weights of the representations $\msR_{j_L}\otimes \msR_{j_R}$ as we continue $\Delta$ in the complex plane.\footnote{It will be important as we progress to take special care of the cases when $\Delta$ is such that $j_{L/R}\in\frac12\mathbb N$; in these cases the weight spaces terminate discontinuously to finite-dimensional representations.} 
In principle we should consider all representations that satisfy \eqref{eq:DHS_CasimirRelation}, so for a given $(\Delta,\mss)$, we also encounter poles associated to highest weight representations arrived at by sending $\Delta\rightarrow\bar\Delta=2-\Delta$ as well as $\mss\rightarrow-\mss$.\footnote{This is consistent with the Lorentzian picture: a spin-$\mss$ field is built out of $\so$ representations labelled by both $(\Delta,\pm\mss)$ while the $(\Delta,\mss)$ representation is isomorphic to that labelled by $(\bar\Delta,-\mss)$ through the $\so$ shadow map \cite{Penedones:2023uqc}.} If we define $(j_L, j_R) = (-\frac{\Delta+\mss}{2}, -\frac{\Delta-\mss}{2})$ then we will denote
\beq 
    (\bar{j}_L, \bar{j}_R) = \left(-\frac{\bar{\Delta}+\mss}{2}, -\frac{\bar{\Delta}-\mss}{2}\right)~,
\eeq
while $\mss\rightarrow-\mss$ is equivalent to $j_L\leftrightarrow j_R$. We will then have pole contributions from any of the representations appearing in
\beq 
    \mathcal{R}_{\Delta,\mss} = \left\{\msR_{j_L}\otimes \msR_{j_R},  \msR_{\bar{j}_L}\otimes \msR_{\bar{j}_R}, \msR_{j_R}\otimes \msR_{j_L},  \msR_{\bar{j}_R}\otimes \msR_{\bar{j}_L}\right\}~.
\eeq
We make a special note that for scalar representations ($j_L=j_R=j$) it is sufficient to consider the smaller set
\beq 
    \mathcal{R}_{\Delta,\text{scalar}} = \left\{\msR_{j}\otimes \msR_{j},  \msR_{\bar{j}}\otimes \msR_{\bar{j}}\right\}
\eeq
as in \cite{Castro:2023dxp}.

The ``mass-shell condition" \eqref{eq:DHS_CasimirRelation} is only a necessary condition to contribute a physical pole to $Z_\text{STT}^2$. Functional determinants come with boundary and regularity conditions on their functional domain and we must impose these on weight spaces satisfying \eqref{eq:DHS_CasimirRelation} to obtain a physical answer. We will state these up-front in an $\mf{su}(2)$ natural language as the following:
\begin{description}
    \item[Condition I. Single-valued solutions:] A configuration constructed from a representation $\msR_L\otimes\msR_R\in\mc R$ must return to itself under parallel transport around any closed cycle in the Euclidean manifold.
    \item[Condition II. Globally regular solutions:] A configuration constructed from a representation $\msR_L\otimes \msR_R\in\mc R$ must be globally regular on the Euclidean manifold. When the base space is homogeneous this means $\msR_L\otimes\msR_R$ lifts from a representation of the {\it isometry  algebra} to a representation of the {\it isometry group}.
\end{description}

Let us first expand upon {\bf Condition I} for spin-$\mss$ fields on $S^3$. A field $\Phi$ living in a representation $\msR_L\otimes \msR_R$ is parallel transported around a cycle, $\gamma$, through the background connections:
\beq
\Phi_f= \msR_L\left[\mc P\exp\left(\oint_\gamma a_L\right)\right]\,\Phi_i\,\msR_R\left[\mc P\exp\left(-\oint_\gamma a_R\right)\right]~.
\eeq
When $a_{L/R}$ are flat this conjugation is trivial. However for the backgrounds appropriate for describing the $S^3$ metric geometry, the background connections take non-trivial holonomies
\beq
\mathcal P\exp\left(\oint_\gamma a_L\right)=u_L^{-1}\,e^{i2\pi L_3\msh_L^{(\gamma)}}\,u_L^{-1}~,\qquad\quad\mathcal P\exp\left(\oint_\gamma a_R\right)=u_R^{-1}\,e^{i2\pi\bar L_3\msh_R^{(\gamma)}}\,u_R^{-1}~,
\eeq
when $\gamma$ wraps one of two lines on the base $S^3$ \cite{Castro:2023dxp}. These lines are Hopf-linked and Wick-rotate to the coordinate positions of the static patch origin and horizon. They yield respective holonomies
\beq\label{eq:SPholos}
\gamma_\text{orig.}\colon\;\;\left(\msh_L,\msh_R\right)=(1,1)~,\qquad \gamma_\text{hor.}\colon\;\;\left(\msh_L,\msh_R\right)=(1,-1)~.
\eeq
More details can be found in App.\,\ref{app:dSCSdict}. The salient point is that a single-valued field will satisfy
\beq 
    \lambda_L \msh_L - \lambda_R \msh_R \in \mathbb{Z}~,
\eeq
for each of the two sets of holonomies in \eqref{eq:SPholos} and for all weights $(\lambda_L,\lambda_R)$ in the representation $\msR_L\otimes\msR_R$.

From cycles wrapping the origin weights must satisfy 
\beq
\lambda_L-\lambda_R\in\mathbb Z
\eeq
to contribute a pole to $Z_\text{STT}^2$. Weights of a highest-weight representation $\msR_{j_{L/R}}$ necessarily take the form 
\beq\label{eq:su2weights}
    \lambda_{L/R}=j_{L/R}-p_{L/R}~,\qquad p_{L/R}\in\mathbb N_0~,
\eeq
simply via the structure of the $\su$ algebra. Single-valuedness around the static patch origin then requires
\beq
j_L-j_R\in\mathbb Z\qquad\Leftrightarrow\qquad\mss\in\mathbb Z~.
\eeq
This condition is the same for all other representations in $\mc R_{\Delta,\mss}$. We pause here to note that while the representation theory in Sec.\,\ref{sec:nonStandardReps} only relies on $j_L-j_R\in\mathbb R$, we now see that only fields with quantized spin can contribute physical poles to $Z_\text{STT}^2$. We will thus fix $\mss\in\mathbb Z$ and consider the analytic structure of $Z^2_\text{STT}$ as a function of $\Delta$. This analytic structure is constrained by single-valuedness around the static patch horizon, which requires
\beq\label{eq:CondI1}
\lambda_L+\lambda_R\in\mathbb Z~.
\eeq
We will return to this condition shortly.

We now address {\bf Condition II}, that configurations contributing to $Z^2_\text{STT}$ are globally regular. Without loss of generality we will state this for $\msR_{j_L}\otimes \msR_{j_R}\in\mc R_{\Delta,\mss}$. For the $S^3$ background in question the isometry group acts transitively. Thus regularity at a point guarantees global regularity as long as the isometry group, $SU(2)_L\times SU(2)_R$, acts faithfully on the field in question: that is, $\msR_{j_L}\otimes\msR_{j_R}$ lifts to a representation of the isometry group. The Peter-Weyl theorem states that these must be finite-dimensional representations of $\su_L\otimes \su_R$  (see, for example, \cite{HoweRoger1992Nha:}), where such representations have weights \eqref{eq:su2weights} satisfying
\beq\label{eq:FiniteDimWeights}
\lambda_{L/R}=j_{L/R}-p_{L/R}~,\qquad0\leq p_{L/R}\leq 2j_{L/R}~, \qquad j_{L/R}\in\frac{1}{2}\mathbb N_0~.
\eeq
To be clear about interpretation: the DHS method instructs us to consider the structure of $Z_\text{STT}^2$ as $\Delta$ continues to the complex plane. The mass-shell condition, \eqref{eq:DHS_CasimirRelation}, then instructs us to consider representations with generically complex highest weights, $j_{L/R}\in\mathbb C$. Such representations are non-standard and infinite-dimensional. However {\bf Condition II} simply tells us that the poles of $Z^2_\text{STT}$ are located at $\Delta\in\mathbb Z_{\leq-\mss}$ and the orders of these poles are correctly counted not by weights of an infinite dimensional representation but instead by \eqref{eq:FiniteDimWeights}. In this counting we notice that weights of finite dimensional representations of $SU(2)$ are centered about zero and so for any weight satisfying $\lambda_L+\lambda_R=N>0$ there is a corresponding weight with $\lambda_L+\lambda_R=-N$. Thus for the purposes of counting the number of weights contributing to a particular pole, we can restate \eqref{eq:CondI1} as
\beq\label{eq:CondI2}
    \lambda_L+\lambda_R=\abs{N}~,\qquad N\in\mathbb Z~
\eeq
as a necessary condition for incorporating {\bf Condition II}.

We pause to note that for minimally coupled scalar fields \eqref{eq:CondI2} is also sufficient to imply {\bf Condition II}. Thus the scalar one-loop determinant can be written as
\beq
    Z_\text{scalar}=\prod_{\overset{(\lambda_L,\lambda_R)}{\;\;\;\;\in\msR_j\otimes\msR_j}}\prod_{N\in\mathbb Z}\left(\abs{N}-\lambda_L-\lambda_R\right)^{-1/2}\times\prod_{\overset{(\bar\lambda_L,\bar\lambda_R)}{\;\;\;\;\in\msR_{\bar j}\otimes\msR_{\bar j}}}\prod_{\bar N\in\mathbb Z}\left(\abs{\bar N}-\bar \lambda_L-\bar \lambda_R\right)^{-1/2}~,
\eeq
where we have written explicitly the product over the two representations appearing in $\mc R_{\Delta,\text{scalar}}$. From here the expression of $Z_\text{scalar}$ as a Wilson spool follows the procedure in \cite{Castro:2023dxp}.

For massive spin-$\mss$ fields, \eqref{eq:CondI2} is no longer sufficient and we must impose additional constraints to reproduce $Z_\text{STT}^2$. For a given $\abs{N}$ in \eqref{eq:CondI2}, {\bf Condition II} additionally implies the weights $\lambda_{L/R}=j_{L/R}-p_{L/R}$ must satisfy
\beq
p_L\geq -\abs{N}-\mss~,\qquad\qquad p_R\geq -\abs{N}+\mss~.
\eeq
While the first of these is always satisfied (for positive $\mss$) the second is an additional constraint on counting the order of the poles appearing in $Z^2_\text{STT}$ and is only non-trivial when $\abs{N}\leq \mss$. For these $2\mss +1$ cases we observe that $\tilde p_R=p_R+\abs{N}-\mss\geq 0$ and \eqref{eq:CondI2} is equivalently written
\beq
j_L+j_R-p_L-\tilde p_R=\mss~,\qquad p_L,\tilde p_R\geq 0~.
\eeq
We can thus treat this as a condition on the weights $(\lambda_L,\tilde\lambda_R)=(j_L-p_L,j_R-\tilde p_R)$ of highest-weight representations $\msR_{j_L}\otimes \msR_{j_R}$ and for each pole arising from this condition being satisfied, it arises $2\mss +1$ times.

Applying this same procedure to all representations appearing in $\mc R_{\Delta,\mss}$ we arrive at
\beq\label{eq:ZsWeightFormula}
    Z_\text{STT}=\prod_{\mathcal{R}_{\Delta,\mss}}
    \prod_{(\lambda_L, \lambda_R)} \left(
        (\mss - \lambda_L - \lambda_R)^{-\frac{2\mss+1}{2}}
        \prod_{\substack{N \in \mathbb{Z}, \\ \abs{N} > \mss} }
        (|N| - \lambda_L - \lambda_R)^{-\frac{1}{2}}
    \right)~,
\eeq 
where the first product is understood to take the product over all pairs $\msR_L\otimes\msR_R\in\mc R_{\Delta,\mss}$ and the second product is taken over all weights $(\lambda_L,\lambda_R)\in\msR_L\otimes\msR_R$ of a particular pair in $\mc R_{\Delta,\mss}$. Where it does not cause confusion we will maintain this shorthand (in both products and sums) for compactness of notation.

As mentioned at the beginning to this section, the local spin-$\mss$ partition function, $Z_{\Delta,\mss}$, includes, in addition to this quasinormal product, the product from integrating over normalizable zero modes, \eqref{eq:Zzero}:
\beq\label{eq:ZisZZ}
    Z_{\Delta,\mss}=Z_\text{zero}Z_\text{STT}~.
\eeq
In the next section we will show how this combination, with the expression of $Z_\text{STT}$ as a product over representation weights, \eqref{eq:ZsWeightFormula}, will lead to the Wilson spool.

\subsection{Constructing the spool}\label{subsect:spoolcons}

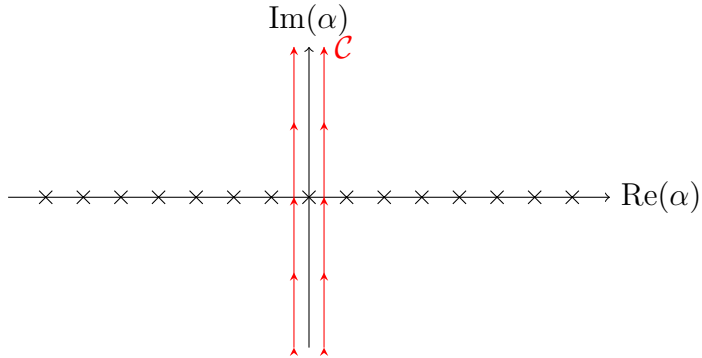
\begin{figure}[h!]
\centering
\begin{tikzpicture}
    \draw[->] (-4, 0) -- (4, 0) node[right] {$\text{Re}(\alpha)$};
    \draw[->] (0, -2) -- (0, 2) node[above] {$\text{Im}(\alpha)$};

    \draw[->-, red] (0.2, -2) -- (0.2, 2) node[right] {$\mathcal{C}$};
    \draw[->-, red] (-0.2, -2) -- (-0.2, 2);

    \foreach \j in {-7,..., 7} {
        \node[cross out, draw=black, inner sep=0pt, outer sep=0pt, minimum size=5] at (0.5*\j, 0) {};
    }
\end{tikzpicture}
\caption{\small The integration contour regulating the $\alpha\rightarrow0$ divergence.\label{fig:contour1}}
\end{figure}

The procedure to cast $\log Z_{\Delta,\mss}$ as an integral over Wilson loop operators starts by rearranging \eqref{eq:Zzero} and \eqref{eq:ZsWeightFormula}. We first make use of the Schwinger parameterization of the logarithm
\beq \label{eq:schwinger}
    \log M = - \int_\times^\infty \frac{\text{d}\alpha}{\alpha} e^{-\alpha M}~,
\eeq
with a regularization of the divergence at $\alpha\rightarrow0$ that we will leave unspecified for now. We will address this regularization through a suitable $i\varepsilon$ prescription below. Applying \eqref{eq:schwinger} to \eqref{eq:ZsWeightFormula}, we first see that the sum over weights in $\log Z_\text{STT}$ can then be organized into representation traces
\beq
    \sum_{(\lambda_L, \lambda_R)} e^{\alpha(\lambda_L + \lambda_R)}
    = \Tr_{\msR_L}\big( e^{\alpha L_3} \big)
    \Tr_{\msR_R}\big( e^{\alpha \bar{L}_3} \big)~,
\eeq
which are the characters of the non-standard representation in \eqref{eq:NSchar}. This then
leads to 
\begin{align}\label{eq:ZSTT-spool}
    \log(Z_\text{STT}) = \frac{1}{2}\int_{\times}^\infty \frac{\text{d}\alpha}{\alpha} 
        \left(\sum_{\substack{N \in \mathbb{Z}\\ \abs{N} > \mss}}e^{-|N|\alpha}
        \right.&+(2\mss+1)e^{-\mss\alpha}
    \left. \vphantom{\sum_{\substack{N \in \mathbb{Z}\\ \abs{N} > \mss}}}\right)
    \times\nonumber\\*
    &\sum_{\mathcal{R}_{\Delta,\mss}}
    \Tr_{\msR_L}\big( e^{\alpha L_3} \big)
    \Tr_{\msR_R}\big( e^{\alpha \bar L_3} \big)~.
\end{align}
Similarly, we can introduce a Schwinger parameter to $\log Z_\text{zero}$, where now  \eqref{eq:Zzero} reads
\begin{align}\label{eq:Zzero-spool}
    \log Z_\text{zero}&=\int_{\times}^\infty \frac{\text{d}\alpha}{\alpha} 
    \left(
        \sum_{n=0}^{\mss-2} ((n+1)^2-\mss^2) e^{(j_L + j_R -n)\alpha}
        -\frac{\mss^2}{2}e^{(j_L + j_R +1)\alpha}
    \right) + (j_{L/R} \to \bar{j}_{L/R}) \nonumber\\
    &=\frac{1}{2}\int_{\times}^\infty \frac{\text{d}\alpha}{\alpha} 
    (e^\alpha - 1)^2 e^{-2\alpha}\left(
        \sum_{n=0}^{\mss-2} ((n+1)^2-\mss^2) e^{-n\alpha}
        -\frac{\mss^2}{2}e^{\alpha}
    \right)\times\nonumber\\
    &\qquad\qquad\qquad\qquad\qquad\qquad\sum_{\mathcal{R}_{\Delta,\mss}}
    \Tr_{\msR_L}\big( e^{\alpha L_3} \big)
    \Tr_{\msR_R}\big( e^{\alpha \bar L_3} \big)~.
\end{align}
In the first line, we used $\Delta=-j_L-j_R$, and in the second, the characters \eqref{eq:NSchar} to cast this as a trace.
The zero mode contribution \eqref{eq:Zzero-spool} combines nicely with $\log\left(Z_\text{STT}\right)$ to give 
\beq 
    \log Z_{\Delta, \mss} = \frac{1}{2}\int_{\times}^\infty \frac{\text{d}\alpha}{\alpha} 
    \left(\frac{\cosh\left(\frac\alpha2\right)}{\sinh\left(\frac\alpha2\right)}
        -\mss^2\sinh(\alpha)
    \right)
    \sum_{\mathcal{R}_{\Delta,\mss}}
    \Tr_{\msR_L}\big( e^{\alpha L_3} \big)
    \Tr_{\msR_R}\big( e^{\alpha \bar L_3} \big)
    ~,
\eeq
where we used
\beq\label{eq:nsum}
    \sum_{n\in\mathbb Z}e^{-\abs{n}\alpha}=\frac{\cosh\left(\frac\alpha2\right)}{\sinh\left(\frac\alpha2\right)}~.
\eeq
At this point we use the even parity of the integrand to regulate the divergence through the following $i\varepsilon$ prescription:
\beq
    \int_{\times}^\infty \frac{\dd\alpha}{\alpha}f(\alpha):=\lim_{\varepsilon\rightarrow 0}\frac{1}{4}\sum_{\pm}\int_{-\infty}^\infty\frac{\dd\alpha}{\alpha\pm i\varepsilon}f(\alpha\pm i\varepsilon)~.
\eeq
This is a choice of regularization scheme for the one-loop determinant. Finally, under a change of integration variables $\alpha\rightarrow-i\alpha$ we can write the partition function as
\begin{align}
    \!\log Z_{\Delta, \mss} = \frac{i}{8}\!\int_{\mathcal{C}} \frac{\text{d}\alpha}{\alpha} 
    \!\left(
        \frac{\cos(\frac{\alpha}{2})}{\sin(\frac{\alpha}{2})}
         + 2\mss^2\cos(\frac{\alpha}{2})\sin(\frac{\alpha}{2})
    \right)\!\sum_{\mathcal{R}_{\Delta,\mss}}
    \Tr_{\msR_L}\big( e^{i \alpha L_3} \big)
    \Tr_{\msR_R}\big( e^{i \alpha \bar L_3} \big)~,
\end{align}
where the contour $\mc C$ runs upwards along the imaginary $\alpha$ axis to the left and right of the divergence at the origin, as depicted in Fig.\,\ref{fig:contour1}.

As a final step we now rewrite the holonomies inside the traces to restore the background connections, arriving at \eqref{eq:logZisW} with $\mathbb W_{j_L,j_R}$ the spinning Wilson spool:
\begin{multline}\label{eq:SpinSpoolDef}
    \mathbb W_{j_L,j_R}[a_L,a_R]:=\frac{i}{2}\int_{\mc C}\frac{\dd\alpha}{\alpha}\frac{\cos\left(\frac\alpha2\right)}{\sin\left(\frac\alpha2\right)}\left(1+2\mss^2\sin^2\left(\frac\alpha2\right)\right)\times\\\sum_{\mc R_{\Delta,\mss}}\Tr_{\msR_L}\left(\mc P e^{\frac{\alpha}{2\pi}\oint_{\gamma}a_L}\right)\Tr_{\msR_R}\left(\mc P e^{-\frac{\alpha}{2\pi}\oint_\gamma a_R}\right)~,
\end{multline}
where above $\gamma=\gamma_\text{hor.}$ is a cycle wrapping the singular point corresponding to the horizon.

At this point let us make several comments:
\begin{itemize}
    \item The spinning Wilson spool takes a form similar to that of the scalar spool found in \cite{Castro:2023dxp,Castro:2023PRL}; importantly the ``operator pieces" of the expression \eqref{eq:SpinSpoolDef} have been organized into gauge-invariant Wilson loop operators. The only modification the spinning spool brings is in the integration measure.
    \item The modification to the integration measure, proportional to $\mss^2$, will have the effect of lowering the degree of each pole at $\alpha\in 2\pi\mathbb Z$ by two. As we will shortly see, this effect reproduces the ``edge partition function" of \cite{Anninos:2020hfj}.
    \item Mathematically the holonomies corresponding to $\gamma_\text{hor.}$ appear in the one-loop determinant because they are sensitive to $\Delta$ on which $Z_\text{STT}$ is treated as an meromorphic function. The physics behind this is clear: we are reproducing a one-loop determinant of massive fields. In the worldline quantum mechanics framework this corresponds to averaging over worldlines of a massive particle in the static patch. Such wordlines are timelike and rotate to a contour gauge equivalent to $\gamma_\text{hor.}$.
\end{itemize}

\subsection{Testing the Wilson spool}\label{sec:SpoolTest}

We now uphold equations \eqref{eq:logZisW} and \eqref{eq:SpinSpoolDef} by verifying that it indeed reproduces the correct path integral of a massive spinning field on $S^3$. There are several ways how to evaluate this path integral, and here we will focus on two approaches to use as a comparison. The first is to implement the DHS method traditionally: in App.\,\ref{app:DHSspinS3} we evaluate this path integral by explicitly listing the quasinormal modes and applying DHS. The second approach we can compare to are the expressions found in \cite{Anninos:2020hfj}, which cast the results in terms of $\so$ characters.

\begin{figure}[h!]
\centering
\begin{tikzpicture}
    \draw[->] (-4, 0) -- (4, 0) node[right] {$\text{Re}(\alpha)$};
    \draw[->] (0, -2) -- (0, 2) node[above] {$\text{Im}(\alpha)$};

    \draw[->-, red] (4,-0.2) -- (0.4, -0.2) arc[start angle=-90, end angle = -270, radius=0.2] --(4, 0.2) node[above] {$\mathcal{C}$};
    \draw[->-, red] (-4,-0.2) -- (-0.4, -0.2) arc[start angle=-90, end angle =90, radius=0.2] --(-4, 0.2);

    \foreach \j in {-7,..., 7} {
        \node[cross out, draw=black, inner sep=0pt, outer sep=0pt, minimum size=5] at (0.5*\j, 0) {};
        }
\end{tikzpicture}
\caption{\label{fig:contour2}{\small We deform the $\alpha$ integration contour to wrap the poles lying along the real axis.}}
\end{figure}
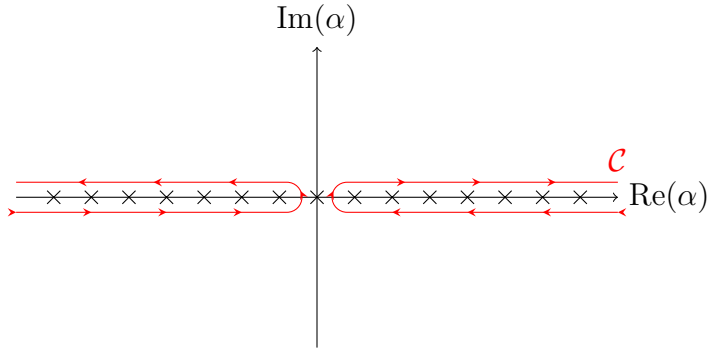

Since we are not turning on gravity ($G_N\to0$),  we evaluate the Wilson loop operators in \eqref{eq:SpinSpoolDef} as characters with the appropriate holonomies in \eqref{eq:SPholos}. Using the form of our non-standard representation character, \eqref{eq:NSchar}, we then write
\beq\label{eq:TLtest1}
    \log Z_{\Delta, \mss} = -\frac{i}{8}\int_{\mathcal{C}} \frac{\text{d}\alpha}{\alpha} 
    \left(
        \frac{\cos(\frac{\alpha}{2})}{\sin^3(\frac{\alpha}{2})}
         + 2\mss^2 \frac{\cos(\frac{\alpha}{2})}{\sin(\frac{\alpha}{2})}
    \right)
    e^{i\alpha(1-\Delta)}~,
\eeq
where we note that the sum over representations in $\mc R_{\Delta,\mss}$ is already neatly packaged into our two contours. We recognize the first term in the parentheses of \eqref{eq:TLtest1} as twice the on-shell scalar Wilson spool found in \cite{Castro:2023dxp}. We evaluate both terms by deforming the $\alpha$ contours to run above and below the real $\alpha$ axis to pick up the residues at the poles lying at $2\pi\mathbb Z_{\neq0}$. This deformation is depicted in Fig.\,\ref{fig:contour2}.

Summing the towers of poles and expressing the $\su_{L/R}$ highest-weight labels in terms of $\mu$ and $\mss$ as in \eqref{eq:HWSpinPS}, we write this as
\begin{align}\label{eq:TreeResult}
\begin{split}
    \log Z_{\Delta,\mss}
    &= \sum_{\pm}\left(
        -\frac{1}{4\pi^2}\Li{3}\left(e^{\mp 2\pi\mu}\right)
        \mp \frac{\mu}{2\pi}\Li{2}\left(e^{\mp 2\pi\mu}\right)
        -\frac{\mu^2+\mss^2}{2}\Li{1}\left(e^{\mp 2\pi\mu}\right)
    \right)\\
    &=2\log Z_{\text{scalar}}-\frac{\mss^2}{2}\sum_{\pm}\Li{1}\left(e^{\mp 2\pi\mu}\right)~,
\end{split}
\end{align}
where
\begin{equation}
    \Li{p}(z) = \sum_{n=1}^{\infty}\frac{z^n}{n^p}
\end{equation}
is the polylogarithm.
As mentioned above, $\log Z_{\text{scalar}}$ is the one-loop determinant of a massive scalar field on $S^3$ with a mass set by $\lds^2m^2=\Delta(2-\Delta)=1+\mu^2$. Up to an overall phase, unfixed by the quasinormal mode method, \eqref{eq:TreeResult} matches the path integral of a massive spin-$\mss$ field on $S^3$  via DHS in App.\,\ref{app:DHSspinS3} and the results reported in \cite{Anninos:2020hfj}.

\section{Coupling quantum matter to \texorpdfstring{dS$_3$}{dS3} quantum gravity}\label{sec:QGcorr}

Having constructed the spinning Wilson spool on $S^3$ in Sec.\,\ref{section:SpoolDerivation}, in this section we will promote it to an off-shell object with the aim of incorporating quantum effects from gravity around $S^3$.
More concretely, we posit the following: let $A_L$ and $A_R$ be $\su_{L}$ and $\su_{R}$  connections respectively, yielding a non-degenerate dreibein $e^a=-i\left(A_L^a-A_R^a\right)$ with an associated metric geometry $(M_3, g_{M_3})$ that is topologically equivalent to the round $S^3$. Then the partition function of massive spin-$\mss$ fields minimally coupled to $g_{M_3}$ is determined by the spinning Wilson spool given by \eqref{eq:SpinSpoolDef} with $a_{L/R}$ replaced with $A_{L/R}$:
\beq\label{eq:OffShellSpool}
\log Z_{\Delta,\mss}[g_{M_3}]=\frac{1}{4}\mathbb W_{j_L,j_R}[A_L,A_R]~.
\eeq

For scalar fields it was argued in \cite{Castro:2023dxp} that the manipulations leading to the Wilson spool remain valid for off-shell geometries. This argumentation relied on the expression of the Laplacian $-\nabla^2_{g_{M_3}}$ as a Casimir of local $\su_{L/R}$ action (see App.\,D of \cite{Castro:2023dxp}). While this remains true when acting on spin-$\mss$ fields (via the transverse-traceless condition), various manipulations leading to \eqref{eq:SpinSpoolDef} make the leap to \eqref{eq:OffShellSpool} less rigorous than for scalar fields. This includes both manipulations relying on the integer nature of the holonomies, $\msh_{L/R}$, as well as the explicit form of $Z_\text{zero}$. Barring a detailed analysis of normalizable zero modes for the ghost and St\"uckelberg fields appearing in $Z_{\Delta,\mss}$ on generic three-geometries, at this point we take \eqref{eq:OffShellSpool} {\it as a proposal}. This proposal is upheld on the grounds that it utilizes and generalizes naturally the gauge-invariant observables in Chern-Simons theory (namely its Wilson loops); these observables appear in a form that reduces to twice\footnote{This stems simply from the fact that a STT field contains roughly two scalars corresponding to the polarizations with $\mss$ and $-\mss$ and has nothing to do with the ambiguities of a compact space and possible zero modes. This counting is obviously not continuous as $\mss\rightarrow0$.} the scalar path integral when $\mss\rightarrow0$ and, as we will see in the following section, continues to work for negative cosmological constant.

With this proposal in hand, the Wilson spool gives us a concrete route for calculating finite $G_N$ effects\footnote{To be clear on scope: here we mean to all orders in $G_N$ perturbation theory about the $S^3$ saddle. We do not consider topology change or other non-perturbative effects here.} to $\log Z_{\Delta,\mss}$. That is we can consider
\begin{align}\label{eq:logZEVisWEV}
    \gev{\log Z_{\Delta,\mss}}:=&\int \left[\mc Dg_{M_3}\right]_{S^3}\,e^{-I_\text{EH}[g_{M_3}]}\log Z_{\Delta,\mss}[g_{M_3}]\nonumber\\
    \equiv& \int [\mc DA_{L}\mc DA_R]_{S^3}\,e^{iS[A_L,A_R]}\left(\frac{1}{4}\mathbb W_{j_L,j_R}[A_L,A_R]\right)~,
\end{align}
where the second line follows from the rewriting of the gravitational variables and action (in a first-order formalism) as two Chern-Simons path integrals.\footnote{In App.\,\ref{app:dSCSdict} we review dictionary between Chern-Simons theory and dS$_3$ gravity.}  Additionally, in the first line, $[\mc Dg_{M_3}]_{S^3}$ indicates that we integrate over metric geometries topologically equivalent to $S^3$, and in the second line, $[\mc DA_L\mc DA_R]_{S^3}$ indicates we perform the Chern-Simons path integral on the base $S^3$ topology.

The power of the second line of \eqref{eq:logZEVisWEV} lies in the breadth of techniques for evaluating Wilson loop observables for $SU(2)$ Chern-Simons theories on the three-sphere \cite{Blau:1993tv,Blau:2006gh,Blau:2013oha,Kapustin:2009kz,Marino:2011nm,Beasley:2005vf,Beasley:2009mb}. In \cite{Castro:2023dxp} it was shown how to adapt two such techniques, Abelianisation \cite{Blau:1993tv,Blau:2006gh,Blau:2013oha} and localization through a $\mc N=2$ supersymmetric extension \cite{Kapustin:2009kz,Marino:2011nm}, for the evaluation of Wilson loop expectation values in light of the features unique to Chern-Simons gravity: non-trivial background connections, complex levels, and non-standard representations appearing in Wilson loop operators. While {\it prima facie} these are two very different techniques, they both lead to the expression of a Wilson loop expectation value as a deformation of its on-shell (i.e. flat) background value integrated over a single modulus:
\beq\label{eq:AbelianizedWEV}
    \int [\mc DA]e^{ik\,S_\text{CS}[A]}\Tr_{\msR}\mc Pe^{\oint_\gamma A}=e^{irS_\text{CS}[a]}\int_{\mathbb R}\dd\sigma\,\sin^2(\pi\sigma)e^{ir\,\frac{\pi}{2}\sigma^2}\chi_{\msR}\left(\sigma+\msh\right)~,
\eeq
where $r=k+2$ is the renormalized level, $a$ is the on-shell background connection, $\msh$ is its holonomy about the contour $\gamma$, and $\chi_{\msR}(z)=\Tr_{\msR}e^{i2\pi z\,L_3}$ is the $\su$ representation character. Equation \eqref{eq:AbelianizedWEV} holds for complex levels and non-standard $\su$ representations, including those found in Sec.\,\ref{sec:nonStandardReps}. Applying this to \eqref{eq:logZEVisWEV}, we can write
\begin{align}\label{eq:AbelianizedlogZEV}
    \!\gev{\log Z_{\Delta,\mss}}=\frac{i}{8}e^{ir_L\SCS[a_L]+ir_R\SCS[a_R]}&\int\dd\sigma_L\dd\sigma_R\, e^{ir_L\frac\pi2\sigma_L^2+ir_R\frac\pi2\sigma_R^2}\sin^2(\pi\sigma_L)\sin^2(\pi\sigma_R)\times\nonumber\\
    &\sum_{\mc R_{\Delta,\mss}}\int_{\mc C}\frac{\dd\alpha}{\alpha}\frac{\cos\left(\frac{\alpha}{2}\right)}{\sin\left(\frac{\alpha}{2}\right)}\left(1+2\mss^2\sin^2\left(\frac{\alpha}{2}\right)\right)\times\nonumber\\
    &\qquad\;\;\chi_{\msR_L}\left(\frac{\alpha}{2\pi}(1+\sigma_L)\right)\chi_{\msR_R}\left(\frac{\alpha}{2\pi}(1-\sigma_R)\right)~,
\end{align}
where we used the off-shell version of the spool in \eqref{eq:SpinSpoolDef}. 
Upon writing
\beq
    r_L=\hat\delta+is~,\qquad r_R=\hat\delta-is~,\qquad s=\frac{\lds}{4G_N}~,\qquad\hat\delta\in\mathbb Z~,
\eeq
per the Chern-Simons gravity dictionary outlined in App.\,\ref{app:dSCSdict}, we posit that \eqref{eq:AbelianizedlogZEV} is {\it exact} (that is, holding to all orders) in $G_N$ perturbation theory about the $S^3$ saddle. 

While this claim has power in principle, in practice \eqref{eq:AbelianizedlogZEV} is complicated as an integral. By rescaling $\sigma_{L/R}\rightarrow r_{L/R}^{-1/2}\sigma_{L/R}$, we can proceed systematically in $\lds^{-1}G_N$ perturbation theory which amounts to a Taylor expansion in $\sigma_{L/R}$ of non-Gaussian pieces of the integrand in $\eqref{eq:AbelianizedlogZEV}$. At any order in this expansion the Gaussian integrals over $\sigma_{L/R}$ can be performed. This leaves the contour integral over $\alpha$ which can be deformed to pick up its poles (which remain at $2\pi\mathbb Z_{\neq0}$ at each order of perturbation theory) in a similar spirit as our computation in Sec.\,\ref{sec:SpoolTest}. This procedure completely mirrors that outlined for the scalar Wilson spool in \cite{Castro:2023dxp} and can be efficiently implemented on a computer algebra system. 

To illustrate this concretely, we evaluate $\gev{\log Z_{\Delta,\mss}}$ normalized by the gravitational path integral to the first non-zero order of $\lds^{-1}G_N$ perturbation theory. The gravitational path integral on $S^3$ has a close form given by
\begin{align}\label{eq:Zgrav}
    \mc Z_\text{grav}=&e^{ir_L\SCS[a_L]+ir_R\SCS[a_R]}\int\dd\sigma_L\dd\sigma_R\,e^{ir_L\frac\pi2\sigma_L^2+ir_R\frac\pi2\sigma_R^2}\sin^2(\pi\sigma_L)\sin^2(\pi\sigma_R)\nonumber\\
    =&e^{2\pi s}\left(ie^{-i2\pi\frac{\hat\delta}{\hat\delta^2+s^2}}\right)\frac{2}{\sqrt{\hat\delta^2+s^2}}\,\abs{\sin\left(\frac{\pi}{\hat\delta+is}\right)}^2~,
\end{align}
as outlined in \cite{Castro:2023dxp} (see also \cite{Castro:2011xb,Anninos:2020hfj,Hikida:2021ese}).
By implementing the rescaling $\sigma_{L/R}\rightarrow r_{L/R}^{-1/2}\sigma_{L/R}$ in \eqref{eq:AbelianizedlogZEV}, to leading order in the coupling we find
\beq
    \frac{\gev{\log Z_{\Delta,\mss}}}{\mc Z_\text{grav}}=\log Z_{\Delta,\mss}[S^3]+\left(\frac{G_N}{\lds}\right)^2\left[\log Z_{\Delta,\mss}\right]_{(2)}+\ldots~,
\eeq
where the dots correspond to subleading corrections in $\lds^{-1}G_N$, and the first non-trivial correction is
\beq
    \left[\log Z_{\Delta,\mss}\right]_{(2)}=\sum_{\pm}\sum_{i=0}^3z_i^{(2)}[\pm \mu,\mss]\Li{-i}\left(e^{\mp 2\pi\mu}\right)~,
\eeq
with
\begin{align}
    z_0^{(2)}[\mu,\mss]=&-\left(\frac{16\pi}{3}-\frac{8}{\pi}-8i\hat\delta\right)\mu^3-\left(16\pi-\frac{144}{\pi}-24i\hat\delta+16\pi\mss^2-\frac{216}{\pi}\mss^2-24i\hat\delta\mss^2\right)\mu\,,\nonumber\\
    z_1^{(2)}[\mu,\mss]=&\left(\frac{16\pi^2}{3}+12-i8\pi\hat\delta\right)\mu^4+\left(16\pi^2-276-i24\pi\hat\delta-432\mss^2\right)\mu^2\nonumber\\
    &-(16\pi^2-372-i24\pi\hat\delta)\mss^4+(16\pi^2-588-i24\pi\hat\delta)\mss^2-48\,,\nonumber\\
    z_2^{(2)}[\mu,\mss]=&-\frac{24\pi}{5}\mu^5+\left(104\pi+96\pi\mss^2\right)\mu^3+\left(96\pi-168\pi\mss^4+792\pi\mss^2\right)\mu\,,\nonumber\\
    z_3^{(2)}[\mu,\mss]=&-\left(8\pi^2-16\pi^2\mss^2\right)\mu^4-\left(32\pi^2+40\pi^2\mss^4+160\pi^2\mss^2\right)\mu^2\nonumber\\
    &+8\pi^2\mss^6+24\pi^2\mss^4-32\pi^2\mss^2~.
\end{align}
We note that taking $\mss\rightarrow0$ doubles the corrections to the scalar one-loop determinant \cite{Castro:2023dxp} as expected. At leading order in a large $\mu$ expansion (which amounts to a large mass expansion while holding $\mss$ fixed)
\beq
    [\log Z_{\Delta,\mss}]_{(2)}=-\frac{48\pi\,\mu^5}{5}e^{-2\pi\mu}+\left(24-\frac{16\pi^2}{3}+32\pi^2\mss^2-i16\pi\hat\delta\right)\mu^4\,e^{-2\pi\mu}+\ldots~,
\eeq
where we have kept to next-to-leading order in the large mass expansion where the first contribution from spin appears.

It is natural at this stage to interpret this as a renormalization of the mass\footnote{Importantly the spin of the field does not get renormalized.} of the spin-$\mss$ field. Writing an expansion for the renormalized mass
\beq\label{eq:mu1}
    \mu_R=\mu+\left(\frac{G_N}{\lds}\right)^2\delta_{\mu}^{(2)}+\ldots~,
\eeq
we note to $O(G_N^2\lds^{-2})$
\beq
\log Z_{\Delta,\mss}=\log Z_{\Delta_R,\mss}-\pi\frac{\cosh(\pi\mu_R)}{\sinh(\pi\mu_R)}(\mu_R^2+\mss^2)\left(\frac{G_N^2}{\lds^2}\right)\delta_\mu^{(2)}+\ldots~,
\eeq
where $\Delta_R=1-i\mu_R$. Interpreting the corrections due to quantum gravity as a mass renormalization then sets as a renormalization condition
\beq\label{eq:MassRenormCond}
    \frac{\gev{\log Z_{\Delta,\mss}}}{\mc Z_\text{grav}}\overset{!}{=} \log Z_{\Delta_R,\mss}~,
\eeq
with $G_N$ held fixed. This then determines the renormalized mass as
\begin{align}\label{eq:massrenorm}
    \delta_\mu^{(2)}=&\frac{1}{\pi}\frac{\tanh(\pi\mu_R)}{\mu_R^2+\mss^2}\left[\log Z_{\Delta,\mss}\right]_{(2)}\nonumber\\
    =&-\frac{48}{5}\mu_R^3\,e^{-2\pi\mu_R}+\left(\frac{24}{\pi}-\frac{16\pi}{3}+32\pi\mss^2-i16\hat\delta\right)\mu_R^2\,e^{-2\pi\mu_R}+\ldots~,
\end{align}
where in the second line we have written the leading and next to leading terms in a large mass expansion. Note that the magnitude of the leading term is $\mss$ independent and consistent\footnote{In \cite{Castro:2023dxp} the renormalization condition was taken (tacitly) as
\beq
    \frac{\gev{\log Z_{\Delta_R,\text{scalar}}}}{\Zg}\overset{!}{=}\log Z_{\Delta,\text{scalar}}~,
\eeq
which leads to an overall minus sign with respect to \eqref{eq:massrenorm}.}with the leading mass renormalization for scalar fields \cite{Castro:2023dxp}.

We note that this is not the only renormalization condition that one may choose to set. We could instead to renormalize Newton's constant, $G_N\rightarrow G_{N,R}$, while holding the mass of the spinning field fixed. To illustrate this, we will set $\hat \delta =0$ and consider the following renormalization condition:
\begin{align}\label{eq:GNRenormCond1}
   \frac{\int[\mc Dg]e^{-I_\text{EH}[g]}Z_{\Delta,\mss}[g]}{Z_{\Delta,\mss}[g_{S^3}]}\Big|_{G_N}=&\mc Z_\text{grav}\left(1+\frac{\gev{\log Z_{\Delta,\mss}}}{\mc Z_\text{grav}}-\log Z_{\Delta,\mss}[g_{S^3}]+\ldots\right)\Big|_{G_N}\nonumber\\
   &\overset{!}{=}\mc Z_\text{grav}\Big|_{G_{N,R}}~,
\end{align}
 or equivalently
\beq\label{eq:GNRenormCond2}
    \frac{\mc Z_\text{grav}\big|_{G_{N,R}}}{\mc Z_\text{grav}\big|_{G_N}}=1+\left(\frac{G_{N,R}}{\lds}\right)^2\left[\log Z_{\Delta,\mss}\right]_{(2)}+\ldots~.
\eeq
Note we have normalized by $Z_{\Delta,\mss}[g_{S^3}]$ in \eqref{eq:GNRenormCond1} as this leading term decouples from metric fluctuations. It would be responsible for a renormalization of the cosmological constant, but does not mediate the gravitational self-interactions relevant for renormalizing the coupling, $G_N$. Writing
\beq
    G_N=G_{N,R}\left(1+\delta_{G_N}^{(2)}\left(\frac{G_{N,R}}{\lds}\right)^2+\ldots\right)~,
\eeq
we find from expanding \eqref{eq:Zgrav} and comparing to the right-hand side of \eqref{eq:GNRenormCond2}
\begin{align}
     \delta^{(2)}_{G_N}=&-\frac{1}{3}\left[\log Z_{\Delta,\mss}\right]_{(2)}\nonumber\\
    =&\frac{16\pi}{5}\mu^5\,e^{-2\pi\mu}-\left(8-\frac{16\pi^2}{9}+\frac{32\pi^2}{3}\mss^2
    \right)\mu^4\,e^{-2\pi\mu}+\ldots~,
\end{align}
where again we have written the leading and next-to-leading terms in a large mass expansion.

It is worth noting that the results of this section, both the renormalization of the field mass, \eqref{eq:massrenorm}, and the renormalization of $G_N$, \eqref{eq:GNRenormCond2}, are novel. To our knowledge, corresponding computations have not been carried out in the metric formulation of dS$_3$.\footnote{In the absence of a cosmological constant, a three-loop computation involving gravitons was done in \cite{Leston:2023ugd}. It seems feasible that one could apply that approach to dS$_3$ and verify \eqref{eq:massrenorm} and \eqref{eq:GNRenormCond2}.} To this end, the results of this section provide concrete and testable predictions of the Chern-Simons formulation of gravity.

\section{Spinning spool on \texorpdfstring{AdS$_3$}{AdS3}}\label{sec:AdSspool}

To illustrate the general utility of the spinning Wilson spool constructed in Sec.\,\ref{subsect:spoolcons}, as well as bolster the {\bf Conditions I \& II} that lead to it in Sec.\,\ref{subsect:su2QMNs}, we can repeat this construction in an AdS$_3$ background. An expression for the one-loop determinant of massive spinning fields on a BTZ black hole as a classical Wilson spool was conjectured in \cite{Castro:2023PRL} based on an extension of the result for a massive scalar field.  In this section we show how to derive the spinning spool on AdS$_3$ from the principles set out in  Sec.\,\ref{subsect:su2QMNs} and demonstrate that the final expression accords with the general result \eqref{eq:MainResult}. 

We wish to compute 
\beq\label{eq:AdSFunDet}
    Z_{\Delta,\mss} = \det\left( -\nabla_{(\mss)}^2 + \lads^2\bar{m}_\mss^2 \right)^{-\frac{1}{2}}
\eeq
with $\nabla^2_{(\mss)}$ being the Laplace-Beltrami operator acting on spin-$\mss$ STT tensors. The background geometry entering in \eqref{eq:AdSFunDet} will be the rotating BTZ black hole \cite{Banados:1992wn,Banados:1992gq} with AdS$_3$ radius $\lads$; properties of this geometry are described in App.\,\ref{app:AdSCSdict}. The effective mass $\lads^2\bar m_\mss^2$ is related to the standard mass, $m^2$, via \cite{Datta:2011za}
\beq
    \lads^2\bar m_\mss^2=\lads^2m^2+\mss(\mss-3)~.
\eeq
We recall that the standard mass is related to the conformal dimension of a dual primary through \cite{Sleight:2017krf}
\beq
    \lads^2 m^2=(\Delta+\mss-2)(\Delta-\mss)~.
\eeq
$Z_{\Delta,\mss}$ is completely captured by this one-loop determinant over STT tensors: in contrast to the previous section, there is no additional zero mode product as AdS$_3$ is non-compact.

The isometry group of Lorentzian AdS$_3$ is $SO(2,2)$ with an algebra isomorphic to $\slalg_L\oplus\slalg_R$ which we will take to be generated by $\{L_0,L_\pm\}$ and $\{\bar L_0,\bar L_\pm\}$, respectively.\footnote{Our conventions for the $\slalg$ algebra follow \cite{Gutperle:2011kf,Castro:2018srf}.}
As done for dS$_3$, it is useful to briefly establish our conventions regarding the description of single-particle states living in AdS$_3$ as representations of $\slalg_L\oplus\slalg_R$. 
Lowest-weight representations $\msR_j^{{\scaleto{\text{LW}}{4pt}}}$ are defined by a lowest-weight state, $|j,0\rangle_{\scaleto{\text{LW}}{4pt}}$, which is annihilated by $L_-$ and labelled by its $L_0$ eigenvalue:
\beq
    L_-|j,0\rangle_{\scaleto{\text{LW}}{4pt}}=0~,\qquad\qquad L_0|j,0\rangle_{\scaleto{\text{LW}}{4pt}}=j|j,0\rangle_{\scaleto{\text{LW}}{4pt}}~.
\eeq
All other states in $\msR^{{\scaleto{\text{LW}}{4pt}}}_j$ are generated by the action of $L_+$ acting on $|j,0\rangle_{\scaleto{\text{LW}}{4pt}}$. This representation has a character
\beq
    \chi_{j,\scaleto{{\scaleto{\text{LW}}{4pt}}}{4pt}}(z)=\mTr_{\msR_j^{{\scaleto{\text{LW}}{4pt}}}}\left(e^{i2\pi zL_0}\right)=\frac{e^{i\pi z(2j-1)}}{2\sinh(-i\pi z)}~.
\eeq
Highest-weight representations $\msR^{{\scaleto{\text{HW}}{4pt}}}_j$ are defined by a highest-weight state, $|j,0\rangle_{\scaleto{\text{HW}}{4pt}}$, which is instead annihilated by $L_+$. We will take by convention its $L_0$ eigenvalue to be $-j$:
\beq
    L_+|j,0\rangle_{\scaleto{\text{HW}}{4pt}}=0~,\qquad\qquad L_0|j,0\rangle_{\scaleto{\text{HW}}{4pt}}=-j|j,0\rangle_{\scaleto{\text{HW}}{4pt}}~.
\eeq
All other states of $\msR^{\scaleto{\text{HW}}{4pt}}_j$ are generated by the action of $L_-$ on $|j,0\rangle_{\scaleto{\text{HW}}{4pt}}$. This representation has a character
\beq
    \chi_{j,{\scaleto{\text{HW}}{4pt}}}(z)=\mTr_{\msR_j^{{\scaleto{\text{HW}}{4pt}}}}\left(e^{i2\pi zL_0}\right)=\frac{e^{-i\pi z(2j-1)}}{2\sinh(i\pi z)}~.
\eeq
Our conventions have been chosen so that the Casimir of both highest and lowest-weight representations is given by 
\beq
    c_2^{\slalg}|j,p\rangle_{\scaleto{\text{LW/HW}}{6pt}}=j(j-1)|j,p\rangle_{\scaleto{\text{LW/HW}}{6pt}}~,\qquad\qquad \forall\;\;|j,p\rangle_{\scaleto{\text{LW/HW}}{6pt}}\in\msR^{{\scaleto{\text{LW/HW}}{6pt}}}_j~.
\eeq

Returning to the functional determinant \eqref{eq:AdSFunDet}, we now want to implement the steps in Sec.\,\ref{subsect:su2QMNs}: the ``mass-shell condition'' and {\bf Conditions I \& II}. To start, we write the Laplace-Beltrami operator in terms of the Casimir of the isometries of AdS$_3$ \cite{Sleight:2017krf}: 
\beq
    2c_2^{\slalg_L}+2c_2^{\slalg_R}=\nabla^2_{(\mss)}+\mss(\mss+1)~.
\eeq
In this language, a pole contributing to $Z_{\Delta,\mss}^2$ corresponds to a state $|\psi\rangle$ in a representation of $\slalg_L\oplus\slalg_R$ satisfying 
\beq
\left(2c_2^{\slalg_L}+2c_2^{\slalg_R}\right)|\psi\rangle=\left(\Delta(\Delta-2)+\mss^2\right)|\psi\rangle~.
\eeq
This is the ``mass-shell condition'' and it is satisfied for pairs of highest and lowest-weight representations labeled by
\beq\label{eq:AdSjLR}
    j_{L}=\frac{\Delta\pm\mss}{2}~,\qquad j_R=\frac{\Delta\mp\mss}{2}~.
\eeq
Note that representations labeled by \eqref{eq:AdSjLR} with $\Delta$ replaced by $\bar\Delta=2-\Delta$ share the same Casimir. However, unlike in dS$_3$, we are forced to choose either $\Delta$ or $\bar\Delta$ due to Dirichlet boundary conditions tacitly imposed on solutions contributing to \eqref{eq:AdSFunDet}.\footnote{Dirichlet boundary conditions also exclude $\slalg$ representations that are neither highest nor lowest-weight from contributing to the one-loop determinant.} For what follows we will assume, without loss of generality, that $\Delta$ corresponds to a normalizable massive spin-$\mss$ solution.\footnote{In the special cases where $\Delta$ and $\bar\Delta$ both correspond to normalizable solutions, we simply choose $\Delta$, again without loss of generality. }  Fixing $j_{L/R}$ as in \eqref{eq:AdSjLR} with the upper sign, we can then have pole contributions from any representation appearing in
\beq
    \mc R_{\Delta,\mss}=\mc R^{{\scaleto{\text{HW}}{4pt}}}_{\Delta,\mss}\cup\mc R^{{\scaleto{\text{LW}}{4pt}}}_{\Delta,\mss}=\left\{\msR_{j_L}^{\scaleto{\text{HW}}{4pt}}\otimes\msR_{j_R}^{\scaleto{\text{HW}}{4pt}},\msR_{j_R}^{\scaleto{\text{HW}}{4pt}}\otimes\msR_{j_L}^{\scaleto{\text{HW}}{4pt}}\right\}\cup\left\{\msR_{j_L}^{\scaleto{\text{LW}}{4pt}}\otimes\msR_{j_R}^{\scaleto{\text{LW}}{4pt}},\msR_{j_R}^{\scaleto{\text{LW}}{4pt}}\otimes\msR_{j_L}^{\scaleto{\text{LW}}{4pt}}\right\}~.
\eeq
For scalars, $\mss=0$, we have the reduced set
\beq
    \mc R_{\Delta,\text{scalar}}=\mc R^{{\scaleto{\text{HW}}{4pt}}}_{\Delta,\text{scalar}}\cup\mc R^{{\scaleto{\text{LW}}{4pt}}}_{\Delta,\text{scalar}}=\left\{\msR_j^{\scaleto{\text{HW}}{4pt}}\otimes \msR_j^{\scaleto{\text{HW}}{4pt}}\right\}\cup\left\{\msR_j^{{\scaleto{\text{LW}}{4pt}}}\otimes\msR_j^{\scaleto{\text{LW}}{4pt}}\right\}~.
\eeq

We now apply the group theoretic conditions of Sec.\,\ref{subsect:su2QMNs} to the representations contributing to \eqref{eq:AdSFunDet}. {\bf Conditions I \& II} as they are stated in Sec.\,\ref{subsect:su2QMNs} can be readily applied to the one-loop determinant on BTZ, and they respectively imply:
\begin{description}
    \item[Condition I. Single-valued solutions.]  In Euclidean signature, BTZ is a solid torus that has two cycles that characterize its global properties. Requiring solutions to be single-valued around the contractible thermal cycle, $\gamma_\text{th}$, of the BTZ geometry requires $\mss\in\mathbb Z$. Requiring solutions to be single-valued around  the non-contractible spatial cycle, $\gamma_\text{sp}$, of the BTZ geometry requires a weight $(\lambda_L,\lambda_R)\in\msR_L\otimes\ms R_R$ to satisfy 
    \beq
    \lambda_L \msh_L - \lambda_R \msh_R \in \mathbb{Z}~,
    \eeq
    where $\msh_{L,R}$ are holonomies around the non-contractible cycle 
    \beq
        \msh_L=-\frac{1}{\tau}~,\qquad\msh_R=-\frac{1}{\bar\tau}~,
    \eeq
    and $\tau$ is the modular parameter defining the geometry. See App.\,\ref{app:AdSCSdict} for details. 
    \item[Condition II. Globally regular solutions.] A representation $\msR_L\otimes\msR_R\in\mc R_{\Delta,\mss}$ is required to lift to a group representation of $SL(2,\mathbb R)\times SL(2,\mathbb R)$~. In contrast to Sec.\,\ref{subsect:su2QMNs}, in the present case {\bf Condition II} is trivial since {\it every} representation of $\slalg$ lifts to a (not necessarily unitary) representation of the universal cover of $SL(2,\mathbb R)$ \cite{kitaev:2018notes}.
\end{description}
Thus we fix $\mss\in\mathbb Z$ and the one-loop determinant \eqref{eq:AdSFunDet} is then the product of poles in the complex $\Delta$ plane given by
\beq\label{eq:AdSZsWeightFormula}
    Z_{\Delta,\mss}=\prod_{\mathcal{R}_{\Delta,\mss}}
    \prod_{(\lambda_L, \lambda_R)}\prod_{N\in\mathbb Z}\left(\abs{N}-\lambda_L\msh_L+\lambda_R\msh_R\right)^{-1/4}\left(\abs{N}+\lambda_L\msh_L-\lambda_R\msh_R\right)^{-1/4}~.
\eeq

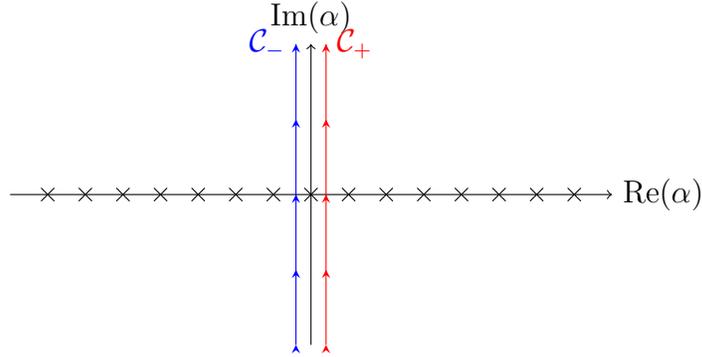
\begin{figure}[h!]
\centering
\begin{tikzpicture}
    \draw[->] (-4, 0) -- (4, 0) node[right] {$\text{Re}(\alpha)$};
    \draw[->] (0, -2) -- (0, 2) node[above] {$\text{Im}(\alpha)$};

    \draw[->-, red] (0.2, -2) -- (0.2, 2) node[right] {$\mathcal{C}_+$};
    \draw[->-, blue] (-0.2, -2) -- (-0.2, 2) node[left] {$\mathcal{C}_-$};

    \foreach \j in {-7,..., 7} {
        \node[cross out, draw=black, inner sep=0pt, outer sep=0pt, minimum size=5] at (0.5*\j, 0) {};
    }
\end{tikzpicture}
\caption{\small{The integration contours relevant for the Wilson spool in AdS$_3$ gravity. Highest-weight representations are integrated against $\mc C_-$ (in blue) while lowest-weight representations are integrated against $\mc C_+$ (in red). The black crosses depict the poles of the integration measure, $\frac{1}{\alpha}\frac{\cos\alpha/2}{\sin\alpha/2}$. Not depicted are the poles of the representation traces which lie along the imaginary $\alpha$ axis.}}\label{fig:AdScontours}
\end{figure}

Following \eqref{eq:schwinger}, we next implement a Schwinger parameterization of the logarithm of this expression which reads
\beq
    \log(Z_{\Delta,\mss}) = \frac{1}{4} \int_\times^\infty \frac{\text{d}\alpha}{\alpha}\frac{\cosh\left(\frac{\alpha}{2}\right)}{\sinh\left(\frac{\alpha}{2}\right)} \sum_{\mathcal{R}_{\Delta,\mss}}\sum_{(\lambda_L, \lambda_R)} \left( e^{\alpha(\lambda_L \msh_L - \lambda_R \msh_R)} +  e^{-\alpha(\lambda_L \msh_L - \lambda_R \msh_R)} \right) ~,
\eeq
where we have performed the sum over $N$ as in \eqref{eq:nsum}. We will regulate the $\alpha\rightarrow0$ divergence of the above expression by combining the two terms in the bracket into a single contour integral regulated about the origin by an $i\varepsilon$ prescription. To ensure convergence of the representation traces, a separate $i\varepsilon$ prescription must be given to the highest and lowest-weight representations appearing in $\mc R_{\Delta,\mss}$:
\beq\label{eq:AdSlogZiep1}
    \log(Z_{\Delta,\mss}) =
    \frac{1}{4} \sum_{\mathcal{R}_{\Delta,\mss}} \int_{-\infty \pm i\varepsilon}^{\infty \pm i\varepsilon} \frac{\text{d}\alpha}{\alpha}\frac{\cosh\left(\frac{\alpha}{2}\right)}{\sinh\left(\frac{\alpha}{2}\right)}\sum_{(\lambda_L, \lambda_R)} e^{\alpha(\lambda_L \msh_L - \lambda_R \msh_R)}~.
\eeq
with the (+/-) sign applying to representations in $\mc R_{\Delta,\mss}^{\scaleto{\text{LW/HW}}{4pt}}$, respectively. Recognizing the sum over weights as a representation trace and redefining $\alpha\rightarrow -i\alpha$ we then can write this suggestively as
\begin{multline}
    \log(Z_{\Delta,\mss}) = \frac{i}{4}
    \sum_{\mathcal{R}^{\scaleto{\text{LW}}{4pt}}_{\Delta,\mss}}\int_{\mathcal{C}_+} 
    \frac{\text{d}\alpha}{\alpha}\frac{\cos\left(\frac{\alpha}{2}\right)}{\sin\left(\frac{\alpha}{2}\right)}
    \Tr_{\msR_L} \left( \mathcal{P} e^{\frac{\alpha}{2\pi} \oint_{\gamma_\text{sp}} a_L} \right)
    \Tr_{\msR_R} \left( \mathcal{P} e^{-\frac{\alpha}{2\pi} \oint_{\gamma_\text{sp}} a_R} \right)\\
    +\frac{i}{4}\sum_{\mathcal{R}^{\scaleto{\text{HW}}{4pt}}_{\Delta,\mss}}\int_{\mathcal{C}_-}\frac{\text{d}\alpha}{\alpha}\frac{\cos\left(\frac{\alpha}{2}\right)}{\sin\left(\frac{\alpha}{2}\right)}
    \Tr_{\msR_L} \left( \mathcal{P} e^{\frac{\alpha}{2\pi} \oint_{\gamma_\text{sp}} a_L} \right)
    \Tr_{\msR_R} \left( \mathcal{P} e^{-\frac{\alpha}{2\pi} \oint_{\gamma_\text{sp}} a_R} \right)~,
\end{multline}
where the contours $\mc C_{\pm}$ are depicted in Fig.\,\ref{fig:AdScontours} and applied separately to the lowest and highest-weight representations appearing in $\mc R_{\Delta,\mss}$.
We have written $\log Z_{\Delta,\mss}$ in this form to draw comparison to the dS$_3$ Wilson spool in Sec.\,\ref{subsect:spoolcons}. In the construction from that section, all representations are integrated along both $\mc C_{\pm}$. This is natural: as was argued in Sec.\,\ref{subsect:su2QMNs}, the poles of the one-loop determinant lie on states of finite-dimensional $\su$ representations which are simultaneously highest and lowest-weight. In the present case, $\mc C_{\pm}$ appear distinctly because highest and lowest-weight representations are distinct for $\slalg$.

With this comparison noted, we can express $\log Z_{\Delta,\mss}$ purely in terms of lowest-weight representations, which are more standard in the AdS/CFT dictionary, by returning to \eqref{eq:AdSlogZiep1} and recalling that every weight of a highest-weight representation is the negative of a weight in a corresponding lowest-weight representation. This allows to write $\log Z_{\Delta,\mss}$ in the form conjectured in the supplemental material of \cite{Castro:2023PRL}
\beq\label{eq:AdSSpinlZfinal}
    \log(Z_{\Delta,\mss}) = \frac{i}{4}
    \sum_{\mathcal{R}_{\Delta,\mss}^{{\scaleto{\text{LW}}{4pt}}}}\int_{2\mathcal{C}_+} 
    \frac{\text{d}\alpha}{\alpha}\frac{\cos\left(\frac{\alpha}{2}\right)}{\sin\left(\frac{\alpha}{2}\right)}
    \Tr_{\msR_L} \left( \mathcal{P} e^{\frac{\alpha}{2\pi} \oint_{\gamma_\text{sp}} a_L} \right)
    \Tr_{\msR_R} \left( \mathcal{P} e^{-\frac{\alpha}{2\pi} \oint_{\gamma_\text{sp}} a_R} \right).
\eeq
providing a principled derivation of that result.  It was shown there that utilizing the holonomies of the background connections corresponding to a spinning BTZ black hole (see App.\,\ref{app:AdSCSdict} for the explicit form), the right-hand side of \eqref{eq:AdSSpinlZfinal} evaluates to the correct one-loop determinant of massive spin-$\mss$ fields on that background:
\beq
    \log Z_{\Delta,\mss}=-\sum_\pm\sum_{l,\bar l=0}^\infty\log\left(1-q^{\frac{\Delta\pm\mss}{2}+l}\bar q^{\frac{\Delta\mp\mss}{2}+\bar l}\right)~,\qquad q=e^{i\frac{2\pi}{\tau}}~,\qquad\bar q=e^{-i\frac{2\pi}{\bar\tau}}~.
\eeq
This agrees with the one-loop determinant of BTZ for spinning fields as reported in, for example, \cite{Giombi:2008vd,David:2009xg,Datta:2011za,Castro:2017}.

We finish this section by reconciling the spinning Wilson spool in AdS$_3$ gravity with our main result \eqref{eq:MainResult}, namely the lack of the $\mss^2$ correction to the $\alpha$ integration measure. This lack stems from two places in our derivation: (i) the absence of normalizable zero-modes and (ii) the triviality of {\bf Condition II} which leaves the product over the weight spaces unrestricted. However it is interesting to note that, on-shell, $\mTr_{\msR_{L/R}}\mc P e^{\pm\frac{\alpha}{2\pi}\oint_{\gamma_\text{sp}}a_{L/R}}$ only have poles along the imaginary $\alpha$ axis \cite{Castro:2023PRL}. Thus this correction term is {\it completely regular} along the real $\alpha$ axis and integrates to zero. We may then include it into the definition of the spinning Wilson spool and state succinctly, for both signs of the cosmological constant,
\beq
    \log Z_{\Delta,\mss}=\frac{1}{4}\mathbb W_{j_L,j_R}[a_L,a_R]~,
\eeq
with $\mathbb W_{j_L,j_R}$ appearing in \eqref{eq:MainResult} and the representations $\msR_{j_{L,R}}$, the integration contour, and the background-connections $a_{L,R}$ chosen appropriately.

\section{Discussion}\label{sec:disc}

In this paper we extended the Wilson spool constructions of \cite{Castro:2023dxp,Castro:2023PRL} to incorporate one-loop determinants of massive spinning fields in both Euclidean dS$_3$ and AdS$_3$ backgrounds. This construction was based upon arranging the quasinormal mode spectra of the respective backgrounds into the representation theory of the isometry algebra.  Along the way we codified two important principles for evaluating one-loop determinants in a representation-theoretic framework. These conditions lead naturally to a spinning Wilson spool expression for the local path integral, $\log Z_{\Delta,\mss}$, of a massive spinning field, \eqref{eq:MainResult}. This expression mimics the scalar Wilson spool: it is an integral over gauge invariant Wilson loop operators with the integral providing a mechanism for ``wrapping" the loops around cycles of the base manifold. The spinning Wilson spool only departs from the scalar expression in its integration measure which cleanly captures the ``edge contributions" of \cite{Anninos:2020hfj}. We posited an off-shell expression for the spinning Wilson spool which allows its insertion into the Chern-Simons path integral. In the context of a Euclidean dS$_3$ background, exact methods in Chern-Simons theory provide efficient methods for evaluating quantum gravity corrections to $\log Z_{\Delta,\mss}$. We discussed two renormalization conditions in which these quantum corrections can be interpreted as renormalizing either the particle mass or Newton's constant. This provides concrete predictions for testing the correspondence between Chern-Simons theory and three-dimensional quantum gravity.

There are multiple comments in order about our results, as well as future directions to which the spinning Wilson spool may prove useful. We briefly discuss these below.

\subsection*{Group theoretic perspective on one-loop determinants}

One important result of Sec.\,\ref{subsect:su2QMNs} is the restating of the quasinormal mode method in a manner directly utilizing aspects of the representation theory and isometries of the background. At the core of that section are {\bf Conditions I \& II} which precisely isolate the representations and the states of a representation that contribute poles to massive one-loop determinants. We regard these conditions to be a significant new asset to the evaluation of one-loop determinants. For one, {\bf Conditions I \& II} will play an important role in establishing the Wilson spool for matter on Euclidean saddles beyond $S^3$, which is ongoing work \cite{wip:lens}. For instance, our preliminary findings indicate that these two conditions correctly predict the placement and the degeneracies of the poles of scalar one-loop determinants on Lens spaces. This gives us strong credence that a Wilson spool built from {\bf Conditions I \& II} will accurately describe the physics of matter on those backgrounds.

More broadly, we have taken care to state the content of these conditions in a manner not specific to spacetime dimension or sign of cosmological constant. Even outside the context of 3d quantum gravity, it is our expectation that {\bf Conditions I \& II} will provide a powerful asset to organizing quasinormal modes and evaluating one-loop determinants for any manifold which possesses a transitive group action, regardless of dimension.

\subsection*{On the role of finite dimensional $\su$ representations}

In Sec.\,\ref{sec:nonStandardReps} we took great care to build $\su$ representations corresponding to massive spinning particle states; these representations are infinite dimensional and non-standard. The reader might then find it surprising that finite dimensional representations played such important role in Sec.\,\ref{subsect:su2QMNs} (particularly in the implementation of {\bf Condition II}). It is worth disentangling the roles of these two separate series of representations.

We regard the mass-shell condition, \eqref{eq:DHS_CasimirRelation}, as always providing a link between representation theory and the particle mass. For physical values of the mass, these representations are generically non-standard. The implementation of {\bf Condition II} is a separate statement about the analytic structure of $Z^{2}_\text{STT}$, namely it encounters poles on finite dimensional representations of $\su$. These however do not (necessarily) lie on physical values of the mass. In evaluating $\mathbb W_{j_L,j_R}$ our goal is not to land on a pole; our goal is to evaluate a one-loop determinant for a physical field. As such the representations appearing in \eqref{eq:MainResult} should be appropriate for a physical value of $\Delta$ and are, again, generically non-standard. Obtaining the right hand side of  \eqref{eq:TreeResult} is a non-trivial verification of this fact. 

We can further illustrate this distinction by investigating what happens if we insert finite dimensional representations $j_{L/R}\in\frac{1}{2}\mathbb Z$ directly into the Wilson spool (say, for the classical $S^3$ background). For the sake of this illustration we will take a non-spinning case $j_L=j_R=j$:
\beq
    \left.\mathbb W_{j}\right|_{\text{finite dim rep}}=\frac{i}{2}\int_{\mc C}\frac{\dd\alpha}{\alpha}\frac{\cos\left(\frac\alpha2\right)}{\sin\left(\frac\alpha2\right)}\frac{\sin^2\left((2j+1)\frac\alpha2\right)}{\sin^2\left(\frac\alpha2\right)}~,
\eeq
where we have inserted directly the finite dimensional $\su$ characters $\chi_j(z)=\frac{\sin\left(\pi(2j+1)z\right)}{\sin(\pi z)}$ and utilized the holonomies around $\gamma_\text{hor.}$, \eqref{eq:SPholos}. However it is clear what will happen if we wrap $\mc C$ as in Fig.\,\ref{fig:contour2}: the poles at $\alpha\in2\pi\mathbb Z_{\neq 0}$ are now only first order and now without any exponential damping. We thus find
\beq\label{eq:finitedim-spool}
    \left.\mathbb W_{j}\right|_{\text{finite dim rep}}=\sum_{n=1}^{\infty}\frac{(2j+1)^2}{n}=(2j+1)^2\zeta(1)=\infty~,
\eeq
where $\zeta$ is the Riemann zeta function. This illustrates the importance of utilizing the non-standard representations (corresponding to physical masses) in $\mathbb W$.

\subsection*{Massive versus massless spinning fields}

It is worth commenting on the important distinction between describing massive spinning fields (the focus of this work) versus massless (higher) spin fields in the Chern-Simons formalism. Our comments are very much motivated by viewing Chern-Simons gravity as an effective field theory. This effective field theory provides a description of the physics below some mass gap; indeed we can expect that massive degrees of freedom can be ``integrated'' out, leaving behind an effective response on the remaining low-energy degrees of freedom. This is precisely what the spinning Wilson spool encapsulates: the response of massive degrees of freedom directly in variables natural to the Chern-Simons path integral.

We contrast this with massless degrees of freedom which cannot be integrated out. Instead their presence must alter the low-energy effective field theory. This is an indication that the one-loop determinants of massless higher-spin fields do not have a Wilson spool description. Instead, we already know they are described by modifying the effective field theory to a theory of {\it higher-spin gravity}. For example, one simple way of describing massless higher-spin fields up to spin-$N$ is  by replacing $SL(2,\mathbb R)\rightarrow SL(N,\mathbb R)$ in asymptotically AdS$_3$ spacetimes \cite{Blencowe:1988gj,Bergshoeff:1989ns,Campoleoni:2010zq} and $SU(2)\rightarrow SU(N)$ in asymptotically Euclidean dS$_3$ spacetimes \cite{Joung:2014qya,Anninos:2020hfj}.

These above comments lead naturally to a line of inquiry: ``How does one couple massive matter to a theory of higher-spin gravity?'' The relations between Wilson lines and particle worldlines in higher-spin gravity have been explored, e.g., in \cite{deBoer:2013vca,Ammon:2013hba}. However we believe the Wilson spool (generalized to, e.g., $SU(N)$ or $SL(N,\mathbb R)$) will be an invaluable tool for a more complete answer to this important open question.

\subsection*{Generalized symmetries in 3d gravity}

Let us offer a final, speculative, remark on the insights that the Wilson spool can possibly lend to our understanding of gravity as an effective field theory. A modern perspective on organizing low-energy physics is the extension of the Landau paradigm of broken symmetries to higher-form and non-invertible (what we will collectively call, generalized) symmetries \cite{Gaiotto:2014kfa,McGreevy:2022oyu}. Attempts to categorize gravity as a low-energy phase through this lens include \cite{Benedetti:2021lxj, Hinterbichler:2022agn, Benini:2022hzx}. An important conjecture along these lines is the wide-held belief that all global symmetries are either gauged or explicitly broken in a UV theory of quantum gravity \cite{Harlow:2018tng} and these statements extend to generalized symmetries.

Chern-Simons theory provides a natural area for exploring these discussions applied to three-dimensional gravity: the Wilson line operators of Chern-Simons theory both generate and are mutually charged under a generalized symmetry.\footnote{We are being imprecise about the distinction between topological operators and topological defects here. Because we often have in mind Euclidean signature, we will call an operator anything that can be inserted into the path integral to obtain an expectation value.} It naturally follows that the Wilson spool is also charged under this generalized symmetry and its non-zero expectation value is a direct sign of the explicit breaking of this symmetry through the inclusion of matter. While the spectrum of operators in Chern-Simons theory might be too large, we note the special role non-standard representation theory plays in this statement: as noted above, the Wilson spool associated to a finite-dimensional $\su$ representation diverges even classically, \eqref{eq:finitedim-spool}. It would be very interesting to make more precise the relation between the Wilson spool and the generalized symmetries of Chern-Simons theory as well as what these relations lend to the question, ``What is 3d gravity?''\footnote{We thank Nabil Iqbal for extended discussions in this direction.}

\section*{Acknowledgements}
We thank Eleanor Harris, Eric D'Hoker, Nabil Iqbal, Per Kraus, Albert Law, Victoria Martin, and Rahul Poddar.  
This work has been partially supported by STFC consolidated grants ST/T000694/1 and ST/X000664/1. JRF is additionally supported by Simons Foundation Award number 620869. JRF thanks the University of Amsterdam for its hospitality. We would like to thank the Isaac Newton Institute for Mathematical Sciences, Cambridge, for support and hospitality during the programme ``Black holes: bridges between number theory and holographic quantum information" where work on this paper was undertaken. This work was supported by EPSRC grant no EP/R014604/1.  This research was supported in part by grant NSF PHY-2309135 to the Kavli Institute for Theoretical Physics (KITP).

\appendix
\numberwithin{equation}{section}

\section{Chern-Simons gravity dictionary}\label{app:conv}

\subsection{Euclidean \texorpdfstring{dS$_3$}{dS3}}\label{app:dSCSdict}

Three-dimensional de Sitter spacetime is a maximally symmetric spacetime with two-sphere spatial slices expanding into future and past infinity. Due to this expansion, inertial observers have access to a finite portion of spacetime called the static patch. A coordinate patch covering one-half of this patch is given by $(t,\rho,\varphi)$ with $\rho\in[0,\frac\pi2)$, $t\in(-\infty,\infty)$, and $\varphi\in[0,2\pi)$ with $\varphi$ a periodic coordinate.  The observer's origin and causal horizon lie at $\rho=0$ and $\rho=\frac\pi2$, respectively. The corresponding metric of the static patch is
\beq\label{eq:LordSmet}
    \frac{\dd s^2}{\lds^2}=-\cos^2\rho\,\dd t^2+\dd\rho^2+\sin^2\rho\,\dd\varphi^2~.
\eeq
Above, the ``de Sitter radius,'' $\lds$, sets the length scale for this maximally symmetric spacetime. For the purpose of moving between the following Chern-Simons description and the metric description, as well for computing one-loop determinants, it will be useful to go to Euclidean signature through the Wick-rotation, $t=-i\tau$. Under this rotation the static-patch rotates to a three-sphere in ``torus coordinates''
\beq\label{eq:EucdSmet}
    \frac{\dd s^2}{\lds^2}=\cos^2\rho\,\dd \tau^2+\dd\rho^2+\sin^2\rho\,\dd\varphi^2~.
\eeq
Absence of a conical singularity at the horizon, $\rho=\frac\pi2$, sets $\tau\sim\tau+2\pi$. The isometry group of this Euclidean space is $SU(2)_L\times SU(2)_R/\mathbb Z_2$ with the $L/R$ denoting left/right group action. We will denote the generators of these two actions as $\{L_a\}_{a=1,2,3}$ and $\{\bar L_a\}_{a=1,2,3}$, respectively.

In accordance with the split structure of this isometry group, we will describe Euclidean dS$_3$ gravity with a pair of $SU(2)$ Chern-Simons theories
\beq\label{eq:dStreeact}
    S=k_L\,\SCS[A_L]+k_R\,\SCS[A_R]~,
\eeq
with
\beq
    \SCS[A]=\frac{1}{4\pi}\mTr\int\left(A\wedge \dd A+\frac{2}{3}A\wedge A\wedge A\right)~,
\eeq
and the trace taken in the fundamental representation. The basic ingredients of the dictionary between this Chern-Simons description and the more familiar ``metric description" is given by relating the gauge connections, $A_{L/R}$, to a dreibein, $e^a$, and (dual) spin-connection, $\omega^a=\frac{1}{2}\epsilon^{abc}\omega_{bc}$, via
\beq
    A_L=i\left(\omega^a+\frac{1}{\lds}e^a\right)L_a~,\qquad\qquad A_R=i\left(\omega^a-\frac{1}{\lds}e^a\right)\bar L_a~.
\eeq
The levels, $k_{L/R}$, are written as
\beq
    k_L=\delta+is~,\qquad k_R=\delta-is~.
\eeq
We can rewrite the action \eqref{eq:dStreeact} as
\beq
    iS=-I_{\text{EH}}-i\delta I_{\text{GCS}}~,
\eeq
with
\beq
    I_{\text{EH}}=-\frac{s}{4\pi\lds}\int\epsilon_{abc}e^a\wedge\left(R^{bc}-\frac{1}{3\lds^2}e^b\wedge e^c\right)
\eeq
the Euclidean Einstein-Hilbert term in the first-order formalism\footnote{Here $R^{ab}$ is the Riemann two-form.} with positive cosmological constant, $\Lambda=\lds^{-2}$. Thus $s\equiv\frac{\lds}{4G_N}$ is the (inverse) gravitational coupling; in this paper we will find it convenient to keep it written as $s$ and keep in mind the classical limit is $s\rightarrow\infty$. $I_{\text{GCS}}$ is the gravitational Chern-Simons term
\beq
    I_\text{GCS}=\frac{1}{2\pi}\mTr\int\left(\omega\wedge \dd\omega+\frac{2}{3}\omega\wedge\omega\wedge\omega\right)+\frac{1}{2\pi\lds^2}\mTr\int e\wedge T~,
\eeq
with $T$ the torsion two-form. Under quantization the levels will undergo a finite renormalization $k_{L/R}\rightarrow r_{L/R}\equiv k_{L/R}+2$ \cite{Witten:1989ip} amounting to a renormalization of the $I_{\text{GCS}}$ coupling, $\delta\rightarrow \hat \delta=\delta+2$. For the rest of this paper we will always work directly with the renormalized levels.

Appropriate classical flat background connections, $a_{L/R}$, describing the $S^3$ are given by
\begin{align}\label{eq:treedSsol}
    a_L=&iL_1\,\dd\rho+i\left(\sin\rho L_2-\cos\rho L_3\right)(\dd\varphi-\dd\tau)~,\nonumber\\
    a_R=&-i\bar L_1\,\dd\rho-i(\sin\rho\bar L_2+\cos\rho\bar L_3)(\dd\varphi+\dd\tau)~.
\end{align}
An important aspect of the above connections is that they each possess ring singularities at $\rho=0$ and $\rho=\frac\pi2$ where the $\varphi$ and $\tau$ coordinates degenerate, respectively. These Wick-rotate to the worldline at the static patch origin and to the causal horizon bifurcation surface, respectively. There is potential for holonomy around these singularities which we will write generically as
\beq
    \mc P\exp\oint_{\gamma}a_{L}=g_\rho^{-1}e^{i2\pi L_3\msh_L}g_\rho~,\qquad\mc P\exp\oint_{\gamma}a_{R}=\bar g_\rho\,e^{i2\pi \bar L_3\msh_R}\bar g_\rho^{-1}~,
\eeq
for periodic group elements, $g_\rho=e^{iL_1\rho}$ and $\bar g_\rho=e^{i\bar L_1\rho}$, and holonomies, $\msh_{L/R}$. Given the solution \eqref{eq:treedSsol}, it is easy to deduce that for cycles, $\gamma_\text{orig}$, wrapping the static-patch origin at $\rho=0$
\beq
    \gamma_\text{orig}\colon\qquad\msh_{L}=1~,\qquad \msh_{R}=1~,
\eeq
while for cycles, $\gamma_\text{hor}$, wrapping the causal horizon at $\rho=\frac\pi2$,
\beq
    \gamma_\text{hor}\colon\qquad\msh_L=1~,\qquad \msh_R=-1~.
\eeq
Lastly we point out that these singularities give delta function sources of curvature\footnote{Importantly the metric geometry remains smooth everywhere.} that is important for reproducing the on-shell action
\beq
    ir_L\SCS[a_L]+ir_R\SCS[a_R]=\frac{\pi\lds}{2G_N}~,
\eeq
which is the tree-level de Sitter entropy.

\subsection{Lorentzian \texorpdfstring{AdS$_3$}{AdS3}}\label{app:AdSCSdict}

For Anti-de Sitter space, we will start directly in Lorentzian signature where the relevant isometry group exhibits a similar split, $SL(2,\mathbb R)_L\times SL(2,\mathbb R)_R$. Accordingly gravity is described by two Chern-Simons theories, this time with opposite levels:
\beq
    S=k \SCS[A_L]-k\SCS[A_R]=S_{\text{EH}}~,\qquad k=\frac{\lads}{4G_N}~,
\eeq
where $S_{\text{EH}}$ is the Lorentzian Einstein-Hilbert action and $\lads$ is the AdS radius. This matching is facilitated by writing the connections as
\beq
    A_L=\left(\omega^a+\frac{1}{\lads}e^a\right)L_a~,\qquad\qquad A_R=\left(\omega^a-\frac{1}{\lads}e^a\right)\bar L_a~,
\eeq
where now $\{L_a\}_{a=0,\pm}$ and $\{\bar L_a\}_{a=0,\pm}$ generate the Lie algebras $\mathfrak{sl}(2,\mathbb R)_L$ and $\mathfrak{sl}(2,\mathbb R)_R$, respectively. We will follow the conventions set in \cite{Gutperle:2011kf,Castro:2018srf} for the $\slalg$ algebra.

The general asymptotically AdS$_3$ vacuum solutions are Ba\~nados geometries \cite{Banados:1998gg} described by background connections
\begin{align}
    a_L=&L_0\,\dd\eta+\left(e^\eta\,L_+-e^{-\eta}\frac{2\pi\mc L}{k}\,L_-\right)\dd x^+~,\nonumber\\
    a_R=&-\bar L_0\,\dd\eta-\left(e^\eta\,\bar L_--e^{-\eta}\frac{2\pi\barmc{L}}{k}\,\bar L_+\right)\dd x^-~,
\end{align}
where $\mc L(x^+)$ and $\barmc L(x^-)$ are arbitrary functions. Here $\eta$ is a radial coordinate with the conformal boudary at $\eta\rightarrow\infty$ and $x^\pm=t\pm \phi$ are the lightcone coordinates of the boundary cylinder. A special case of interest in this paper are the rotating BTZ black hole geometries described by
\beq
    \mc L=\frac{M\,\lads+J}{4\pi}~,\qquad\qquad \barmc{L}=\frac{M\,\lads-J}{4\pi}~,
\eeq
where $M$ and $J$ are the mass and spin of the black hole, respectively \cite{Banados:1992gq,Banados:1992wn}.

In Euclidean signature, $t\rightarrow it_E$, smoothness of the horizon requires that Euclidean time is compact and so the geometry is the interior of a complex torus
\beq
    (w,\bar w)\sim(w+2\pi m+2\pi n\tau,\bar w+2\pi m+2\pi\bar\tau)~,\qquad m,n\in\mathbb Z~,
\eeq
where $w=\phi+it_E$ and $\bar w=\phi-it_E$ and with modular parameter
\beq
    \tau=\frac{i}{2}\sqrt{\frac{k}{2\pi\mc L}}~,\qquad\qquad \bar\tau=-\frac{i}{2}\sqrt{\frac{k}{2\pi\barmc L}}~.
\eeq
Around the two cycles of this torus the connections have holonomy
\beq
    \mc P\exp\oint_\gamma a_L=u_L^{-1}\,e^{i2\pi L_0\msh_L}\,u_L~,\qquad\qquad \mc P\exp\oint_\gamma a_R=u_R\,e^{i2\pi\bar L_0\msh_R}\,u_R^{-1}~,
\eeq
where $u_{L/R}$ are periodic along both cycles (only depending on $\eta$). For the contractible, thermal, cycle $\gamma_\text{th}:(w,\bar w)\rightarrow (w+\tau,\bar w+\bar\tau)$ the connections have holonomy
\beq
    \gamma_\text{th}:\qquad \msh_L=-1~,\qquad \msh_R=-1~,
\eeq
while for the non-contractible spatial cycle, $\gamma_\text{sp}:(w,\bar w)\rightarrow (w+2\pi,\bar w+2\pi)$, they have holonomy
\beq
    \gamma_\text{sp}:\qquad \msh_L=-\frac{1}{\tau}~,\qquad\msh_R=-\frac{1}{\bar\tau}~.
\eeq

\section{DHS for massive spinning fields on \texorpdfstring{$S^3$}{S3}}\label{app:DHSspinS3}

In this appendix, we will review DHS \cite{Denef:2009kn} as it was originally intended: first, determine the quasinormal modes from the equations of motion of a massive spinning field, Wick-rotating them carefully to Euclidean space, and then evaluate the one-loop as a product of the poles. 

Special care is needed when implementing DHS for massive spinning fields. The quasinormal modes of massive fields on de Sitter are found directly from the equations of motion of the field in the manner of \cite{Lopez_Ortega:2006}. However, as shown in \cite{Grewal:2022hlo}, for spin $\mss>0$  some care is required in Wick-rotating de Sitter quasinormal modes into regular modes on $S^3$, that is symmetric, transverse, traceless spherical harmonics. In particular, imposing the na\"ive fall-off condition on the frequency $\omega$ is not sufficient. We will review carefully and explicitly this regularity condition.

We will then construct the one-loop determinant from the quasinormal mode frequencies of a spin-$\mss$ field on $S^3$ using the DHS method and regularise it via zeta function regularisation, as in the case for a scalar field in \cite{Denef:2009kn}. Identities of special functions are then utilised to rewrite these expressions in the form of polylogarithms to make explicit the comparison to the one-loop determinants derived in \cite{Anninos:2020hfj} and Sec.\,\ref{sec:SpoolTest}.

The metric for the de Sitter static patch in ``torus coordinates" is given by \eqref{eq:LordSmet}. For the sake of finding the quasinormal modes of this background, we will also make use of the radial coordinate
\beq
r = \sin\rho~.
\eeq

\subsection{Real massive scalar field}

To introduce notation and the basic ingredients needed in subsequent sections, we will start by deriving the quasinormal modes of a real scalar field in dS$_3$ and then review the derivation of the one-loop determinant as done originally in DHS \cite{Denef:2009kn}. This will set up the procedure for massive spinning fields in the next subsections. 

\subsubsection*{Quasinormal modes}

First, we construct the solutions to the equations of motion for a scalar field $\Phi$ on dS$_3$ 
\beq
    \nabla^2 \Phi = \lds^2 m^2 \Phi~,
\eeq
where $\nabla^2$ is the Laplacian on \eqref{eq:LordSmet} and as in the main text $ \lds^2 m^2 = \Delta (2-\Delta)$. 
Suitable boundary conditions must also be imposed on the solutions to ensure that they Wick-rotate to a regular solution on $S^3$. These are:
\begin{itemize}
    \item The solution must be single-valued in $\varphi$. 

    \item  The solution must be single-valued in imaginary time.

    \item The solution must be regular at the origin of the static patch $(r=0)$.

    \item The solution must be regular and purely in-going or out-going at the horizon $(r=1)$.
\end{itemize}
Proceeding via separation of variables we encounter a second order ODE for $r$, resulting in two solutions. Imposing regularity at the origin removes one of these solutions leaving us with the modes
\beq 
    \Phi = 
    e^{-i\omega t + il\varphi} \left(r^2\right)^{|\frac{l}{2}|} 
    \left(1-r^2\right)^{-\frac{i\omega}{2}}
    \hypgeo{a}{b}{c}{r^2}~,
\eeq
where 
\beq
    a=\frac{\Delta + |l| - i\omega}{2}~, 
    \qquad b=\frac{\bar{\Delta} + |l| - i\omega}{2}~, 
    \qquad c=1+|l|~,
\eeq
and $l, i\omega\in\mathbb{Z}$ are constants constrained to the integers by the single-valuedness of $\Phi$ around $\varphi$ and imaginary time respectively.

To consider regularity at the horizon we will expand the solution to lowest order as $\frac{\varepsilon^2}{2}=1-r^2$ approaches zero.
Taking care as we are in a non-generic case for the hypergeometric function when $c-a-b=i\omega\in\mathbb Z$,
the connection formulae for the hypergeometric function about $r^2$ and $1-r^2$ \cite{Wang_Guo_1989} imply that a purely ingoing or outgoing solution will require at least one of $a,b,(c-a),(c-b)\in \mathbb{Z}_{\leq 0}$. 
Noting that $a\to b$ simply takes $\Delta\to\bar{\Delta}$ which has no effect on the form of the mode $\Phi$, we need consider only two (not necessarily distinct) cases:
\begin{description}
    \item[Case 1:] $a = -n \in \mathbb{Z}_{\leq 0}\implies \Delta = -2n -|l| + i\omega$.
    \item[Case 2:] $c-b = -n \in \mathbb{Z}_{\leq 0}\implies \Delta = -2n -|l| - i\omega$. 
\end{description}
In both cases, the hypergeometric function takes the form of a Jacobi polynomial in terms of Pochhammer symbols (for the rising factorial). For \textbf{Case 1}, $a = -n$ gives
\beq\label{eq:ScalarNearHorizonLimit}
    \hypgeo{-n}{b}{c}{1-\frac{\varepsilon^2}{2}} = \sum_{d=0}^n \left(\frac{\varepsilon^2}{2}\right)^d \frac{(b)_d (c-b)_{n-d}}{(c)_n}~,
\eeq
and for \textbf{Case 2} we make use of the identity $\hypgeo{a}{b}{c}{u} = (1-u)^{c-a-b}\hypgeo{c-b}{c-a}{c}{u}$. Substituting for $a,b,c$, we then find that to lowest order in $\varepsilon$ the following: 
\begin{description}
    \item[Case 1]: $\Delta = -2n -|l| + i\omega~, \qquad n\in \mathbb{Z}_{\leq 0}, i\omega\in\mathbb{Z}~,$
    \begin{equation}
        \Phi \sim \begin{cases}
            e^{-i\omega t} \varepsilon^{+i\omega} \qquad\text{(anti-quasinormal)} & 0 \leq i\omega  \leq n ~,\\
            e^{-i\omega t} \varepsilon^{-i\omega} \qquad\text{(quasinormal)} & \text{else}~,
        \end{cases}
    \end{equation}
    \item[Case 2]: $\Delta = -2n -|l| - i\omega~, \qquad n\in \mathbb{Z}_{\leq 0}, i\omega\in\mathbb{Z}~,$
    \begin{equation}
        \Phi \sim \begin{cases}
            e^{-i\omega t} \varepsilon^{-i\omega} \qquad\text{(quasinormal)} & 0 \leq -i\omega  \leq n ~,\\
            e^{-i\omega t} \varepsilon^{+i\omega} \qquad\text{(anti-quasinormal)} & \text{else}~.
        \end{cases}
    \end{equation}
\end{description}
For quasinormal modes regularity demands $i\omega \leq 0$, for anti-quasinormal modes regularity demands $i\omega \geq 0$. Note that for $i\omega=0$ the modes are equivalent.

For $i\omega\geq0$ \textbf{Case 2} is a subset of \textbf{Case 1} and the possible regular anti-quasinormal modes are 
\beq
\begin{alignedat}{3}
    \Delta &= -2\tilde{n} -|l| - i\omega~, && \qquad n-i\omega\equiv\tilde{n}\in\mathbb{N}_0~.
\end{alignedat}
\eeq
For $i\omega<0$ \textbf{Case 1} is a subset of \textbf{Case 2} and the possible regular quasinormal modes are
\beq
\begin{alignedat}{3}
    \Delta &= -2\tilde{n} -|l| + i\omega~, && \qquad n+i\omega\equiv\tilde{n}\in\mathbb{N}_0~.
\end{alignedat}
\eeq
These two expressions can be combined by allowing $i\omega\equiv k\in\mathbb{Z}$ to range over all integers (and relabelling $\tilde{n}\to n$ for convenience). The Euclidean regular (anti-)quasinormal modes take the values
\beq\label{eq:scalarEuclideanModes}
    \Delta = -2n -|l| - |k|~.
\eeq
Note that the modes with $\Delta \leftrightarrow \bar{\Delta}$ are equivalent due 
to the ambiguity in defining $\Delta$ from the mass. This was seen above by the invariance of $\Phi$ to the exchange, and we have hence counted these modes only once. In using the DHS method we consider the one-loop determinant as a meromorphic function of $\Delta$. Since the determinant also cannot distinguish between $\Delta \leftrightarrow \bar{\Delta}$ there must be a pole at both $\Delta$ and $\bar{\Delta}$. The poles in the one-loop determinant thus appear at 
\beq
    \Delta, \bar{\Delta} = -2n -|l| - |k|~.
\eeq
It can also be confirmed that the modes we have obtained are (expectedly) the scalar spherical harmonics on $S^3$. More simply the degeneracy of each value of $\Delta=-p, p \in \mathbb{Z}_{\geq0}$ can be counted as $(p+1)^2$ matching the degeneracy of the spherical harmonics fitting into representations of $\mathfrak{so}(4)$.

\subsubsection*{One-loop determinant}

Following \cite{Denef:2009kn} the one-loop determinant can be constructed from the location of the quasinormal frequencies, treating the functional determinant as a meromorphic function of $\Delta$. For convenience in this section the notation of \cite{Denef:2009kn} is also followed such that $\Delta\equiv\Delta_+$, and  $\bar{\Delta}\equiv\Delta_-$.

Up to a holomorphic factor, which in this case is trivial, the partition function is simply a product of poles at the (anti-)quasinormal mode frequencies:
\beq\label{eq:Zs-app}
    Z_{\text{scalar}} = \prod_{\substack{\pm, n\in\mathbb{N} \\ l,k \in \mathbb{Z}}} 
    \left(
        |k|+2n + |l| + \Delta_{\pm}
    \right)^{-\frac{1}{2}}~.
\eeq
Noting that $|k|+|l| + 2n=p$ has degeneracy $(p+1)^2$, we can rewrite \eqref{eq:Zs-app} as
\begin{align}
\begin{split}\label{eq:logZs-app}
    \log(\Zs)
    &= -\frac{1}{2} \sum_{\pm, p\in\mathbb{N}} (p+1)^2 
    \log\left(p + \Delta_{\pm}\right) \\
    &= \frac{1}{2} \sum_{\pm} \zeta'(-2,1 \pm i\mu) \mp 2i\mu \zeta'(-1,1 \pm i\mu)
    -\mu^2 \zeta'(0,1 \pm i\mu)~,
\end{split}
\end{align}
where  $\Delta_\pm = 1 \pm i\mu$ for the principle series and $\zeta$ is the Hurwitz zeta function. This is the formula derived in \cite{Denef:2009kn} and it will be convenient to write in terms of polylogarithms. 
The two key the identities are
\begin{align}
\begin{split}
    &\zeta'(-2,\pm i\mu)\mp 2i\mu \zeta'(-1,\pm i\mu)-\mu^2 \zeta'(0,\pm i\mu)   \\
    = &\zeta'(-2,1\pm i\mu)\mp 2i\mu \zeta'(-1,1\pm i\mu) -\mu^2 \zeta'(0,1\pm i\mu)~, 
\end{split}
\end{align}
and\footnote{The identity \eqref{eq:zeta2} can be derived by adapting the proof from \cite{Adamchik_1997} to the region of the complex plane where $x\in i\mathbb{R}$.}
\beq \label{eq:zeta2}
    \zeta'(-n,x) + (-1)^n \zeta'(-n,1-x) = 
    -\pi i \frac{B_{n+1}(x)}{n+1} + e^{\frac{\pi i n}{2}}\frac{n!}{(2\pi)^n}
    \text{Li}_{n+1}(e^{-2\pi i x})~,
\eeq
where $B_n$ are the Bernoulli polynomials. Substituting these into \eqref{eq:logZs-app} we find 
\beq \label{eq:logZs-app-1}
    \log(\Zs) = \sum_{\pm}\left(
    -\frac{\mu^2}{4} \text{Li}_1(e^{\pm 2\pi \mu})
    \pm \frac{\mu}{4\pi}\text{Li}_2(e^{\pm 2\pi \mu})
    - \frac{1}{8\pi^2}\text{Li}_3(e^{\pm 2\pi \mu})\right)~,
\eeq 
 in agreement with \cite{Anninos:2020hfj}.
 
\subsection{Massive spin-1 field}

Next we work out in detail the quasinormal modes of a massive spin-1 field on dS$_3$, and then its one-loop determinant via DHS. Many results follow from the analysis of a scalar field. However, it is important to highlight the subtleties of the regularity condition in this case, and how it reproduces the appearance of the ``edge'' partition function  \cite{Grewal:2022hlo}.

\subsubsection*{Quasinormal modes}

The equation of motion is
\begin{equation}\label{eq:spin1-eom}
\left(\nabla^2_{(1)} -\lds^2m^2 -2\right)\Phi_\mu = 0~,
\end{equation}
accompanied by a transversality condition
\begin{equation}\label{eq:transversalityCondition}
    \nabla^\mu \Phi_\mu = 0~.
\end{equation}
These equations of motion are solved in \cite{Castro:2017} for an AdS$_3$ background using a first-order formalism, we present here the corresponding results in dS$_3$.\footnote{One method to derive these solutions is the analytic continuation of $\lds \to i\lads$ at which point the results of \cite{Castro:2017} can be used directly.} 

 As in the scalar case we discard one set of solutions to the radial part of the PDE as non-regular at the origin, and impose $l, i\omega \in \mathbb{Z}$ to ensure single-valuedness. The resulting expressions are
\begin{align}\label{eq:Spin1QNM_Functions}
\begin{split}
   \Phi_L &= e_L
    e^{-i\omega t + il\varphi} \left({r^2}\right)^{|\frac{l}{2}|} 
    \left(1-r^2\right)^{-\frac{i\omega}{2}}
    \hypgeo{\frac{\bar{\Delta} +1 + |l| - i\omega}{2}}
    {\frac{\Delta -1 + |l| - i\omega}{2}}{1+|l|}{r^2}~, \\
    \Phi_R &= -e_R
    e^{-i\omega t + il\varphi} \left(r^2\right)^{|\frac{l}{2}|} 
    \left(1-r^2\right)^{-\frac{i\omega}{2}}
    \hypgeo{\frac{\bar{\Delta} -1 + |l| - i\omega}{2}}
    {\frac{\Delta +1 + |l| - i\omega}{2}}{1+|l|}{r^2}~, \\
    \Phi_r &= -\frac{i}{(\Delta - 1)} 
    \frac{1}{r(1-r^2)}
    \left[
        (\omega + il)\Phi_L  + (\omega - il)\Phi_R 
    \right]~,
\end{split}
\end{align}
where $\Phi_{L/R}$ are related to the components of the vector field via
\begin{align}
\begin{split}
    i\Phi_\varphi &= \Phi_L + \Phi_R~,\\
    \Phi_t &= \Phi_L - \Phi_R~.
\end{split}
\end{align}
 In \eqref{eq:Spin1QNM_Functions} $e_{L/R}$ are constants of integration and describe the polarisation vector. These constants are further constrained by \eqref{eq:spin1-eom}, \eqref{eq:transversalityCondition} to satisfy one of two conditions depending on the sign of $l$.
\begin{align}\label{eq:Spin1_ModeConditions}
\begin{split}
       (-\omega+il + i(\Delta-1))  e_R  &= (-\omega-il + i(\Delta-1))e_L \,, \qquad l \leq 0~, \\
         (-\omega+il - i(\Delta-1)) e_R &= (-\omega-il - i(\Delta-1))e_L \,,  \qquad l \geq 0~.
\end{split}
\end{align}

Next, we look for those modes which Wick-rotate into regular Euclidean modes on $S^3$. In order for a mode to be single-valued and regular a necessary condition is for $\Phi_L, \Phi_R$ themselves to be single-valued and regular at the horizon. Applying the results from the scalar case \eqref{eq:scalarEuclideanModes} to \eqref{eq:Spin1QNM_Functions} with $\Delta \to \Delta \pm 1$ implies that regular modes potentially occur at
\begin{alignat}{3}
    \Phi_L: \qquad\Delta &= 1 -2N_L -|l| - |k| \label{eq:Spin1_RegularModeConditionsVL}\\
        \text{or }\bar{\Delta} &= 1 -2(N_L'+1) -|l| - |k| \nonumber\\
    \Phi_R: \qquad\Delta &= 1 -2(N_R'+1) -|l| - |k| \label{eq:Spin1_RegularModeConditionsVR}\\
         \text{or }\bar{\Delta} &= 1 -2N_R -|l| - |k|~. \nonumber
\end{alignat}
Unlike in the scalar case there is no symmetry of $\Delta \to \bar{\Delta}$ in $\Phi_\mu$ and so the two cases result in distinct physical modes. 

For $\Phi_\mu$ to have a mode at some value of $\Delta$ or $\bar{\Delta}$ it is necessary for both $\Phi_L$ and $\Phi_R$ to support a mode. We will suppose that $k,l$ are fixed and consider two\footnote{There is also a single exceptional case where $\Delta=\bar{\Delta}=1$ and we can choose a mode in each component by taking $N_L = N_R = l = k = 0$. This also automatically satisfies the constraints of \eqref{eq:Spin1_ModeConditions}. It can be checked using the regularity conditions below that this solution cannot be regular at both the origin and horizon.} families of solutions: 
\begin{itemize}
\item A mode with $N_{L/R}\geq1$ in one of the two components and $N_{R/L}' = N_{L/R}-1 \geq 0$ in the other. We denote these the `general modes' valid for all $N'\geq0$:
\begin{center}\label{tab:Spin1_GeneralModes}
\begin{tabular}{c | c}
    values & ranges \\
    \hline 
    \\[-1em]
    $\Delta = -1 -2N' -|l|-|k|$ & $k,l \in \mathbb{Z}, N' \in \mathbb{Z}_{\geq0}$ \\
    \hline 
    \\[-1em]
    $\bar{\Delta} = -1 -2N' -|l|-|k|$ & $k,l \in \mathbb{Z}, N' \in \mathbb{Z}_{\geq0}$ \\
\end{tabular}
\end{center}
\item If $N_{L/R}=0$ in one component it may be possible to still have a consistent mode if the other component is identically $0$. For example setting $N_{L}=0$ in \eqref{eq:Spin1_RegularModeConditionsVL} it may be permissible from \eqref{eq:Spin1_ModeConditions} to set $e_R = 0$.
A unique list (accounting for overlaps where $k,l=0$) of these `special modes' is
\begin{center}\label{tab:Spin1_SpecialModes}
\begin{tabular}{c | c}
    values & ranges \\
    \hline 
      & $k<0, l\geq 0$ \\
     $\Delta = 1 -|l|-|k|$ & $k=0, l \in \mathbb{Z}$ \\
      & $k>0, l\leq 0$ \\
    \hline 
      & $k<0, l\leq 0$ \\
     $\bar{\Delta} = 1 -|l|-|k|$ & $k=0, l \in \mathbb{Z} \setminus \{0\}$ \\
      & $k>0, l\geq 0$ \\
\end{tabular}
\end{center}
\end{itemize}

Since the regularity of $\Phi_L,\Phi_R$ is only a necessary condition it is still important to check the regularity of the Euclidean norm $|\Phi|^2 = \Phi_\mu \Phi^\mu$. This will determine regularity in a coordinate-independent manner. Consider first regularity as $r\to0$ at the origin of the static patch. The only cases which can cause divergences are those for which $l=0$. In this case, to lowest order in $r$ we find
\beq
    |\Phi|^2 \sim \frac{1}{r^2}\left(
        \frac{|\omega|^2}{(\Delta-1)^2} + 1
    \right)\left| e_L - e_R \right|^2~.
\eeq
The first bracket is always non-zero due to the reality of $\Delta$ and so
regularity implies $e_L =e_R$. For $l=0$ general 
modes this is always true from \eqref{eq:Spin1_ModeConditions}. For the special modes one of $e_{L/R}$ is zero 
and the other is non-zero and so these modes are irregular. 

As in the scalar case, looking at the horizon we can expand in small $\varepsilon$ for $r^2=1-\frac{\varepsilon^2}{2}$. Again using the expansion
\eqref{eq:ScalarNearHorizonLimit} the only potentially 
problematic case is $k=0$, in which case this expansion gives to leading order
\begin{equation}\label{spin 1 dS  horizon regularity condition}
    |\Phi|^2 \sim \frac{1}{\varepsilon^2}
    \left( 1 + \frac{l^2}{(\Delta-1)^2} \right)
    \left|
        e_L \frac{(c_L - b_L)_{n_L}}{(c_L)_{n_L}} + 
        e_R \frac{(c_R - b_R)_{n_R}}{(c_R)_{n_R}}
    \right|^2~,
\end{equation}
where again the constants $a=-n,b,c$ are defined as coefficients in the hypergeometric function for each component.
The general modes with $k=0$ satisfy 
\begin{equation}
    \Delta = -1 - 2N' -|l| \qquad \text{ or } \qquad \bar{\Delta} = -1 - 2N' -|l|~.
\end{equation}
Inserting the first of these into the coefficients of the hypergeometric function
and noting that from \eqref{eq:Spin1_ModeConditions} it is possible to choose
\begin{align}
\begin{split}
    e_L &= +|l|-\Delta+1 = 2(N' + 1 + |l|)~, \\
    e_R &= -|l|-\Delta+1 = 2(N' + 1)~,
\end{split}
\end{align}
it therefore follows\footnote{The identities $(-n)_n = (-1)^n n!$ and 
$(1+|l|)_{N'+1} = (1+|l|)_{N'}(1+|l|+N')$ are useful here.} that
\begin{align}
\begin{split}
    &e_L \frac{(c_L - b_L)_{n_L}}{(c_L)_{n_L}} + 
    e_R \frac{(c_R - b_R)_{n_R}}{(c_R)_{n_R}} \\
    =& 2(N'+1+|l|)\frac{(-N'-1)_{N'+1}}{(1+|l|)_{N'+1}} +
    2(N' + 1 )\frac{(-N')_{N'}}{(1+|l|)_{N'}} = 0~.
\end{split}
\end{align}
The same process works in the $\bar{\Delta}$ case. All the general modes therefore are regular.

In the special cases with $k=0$ and $\Delta, \bar{\Delta} = 1 - |l|$ either $\Phi_L=0,\Phi_R\neq0,n_R=0$ or $\Phi_R=0,\Phi_L\neq0,n_L=0$. Noting that $(x)_0\equiv 1$ the leading order term does not vanish and the special modes with $k=0$ are not regular. Combining the results of regularity at the origin and horizon the final list of regular Euclidean modes is 
\begin{center}
\begin{tabular}{c | c}
    values & ranges \\
    \hline 
    \\[-1em]
    $\Delta = -1 -2N' -|l|-|k|$ & $k,l \in \mathbb{Z}, N' \in \mathbb{Z}_{\geq0}$ \\
    \hline 
    \\[-1em]
    $\bar{\Delta} = -1 -2N' -|l|-|k|$ & $k,l \in \mathbb{Z}, N' \in \mathbb{Z}_{\geq0}$ \\
    \hline 
        & $k<0, l> 0$ \\
        $\Delta = 1 -|l|-|k|$ &  \\
        & $k>0, l< 0$ \\
    \hline 
        & $k<0, l< 0$ \\
        $\bar{\Delta} = 1 -|l|-|k|$ &  \\
        & $k>0, l> 0$ \\
\end{tabular}
\end{center}
It can be confirmed that the degeneracy of the modes at $\Delta=-p, p\geq 1$ is $p(p+2)$. This aligns with the expected degeneracies of transverse spherical vector harmonics on $S^3$, which fit into representations of $\mathfrak{so}(4)$.

As in the scalar case we note that while each of these modes is distinct, the one-loop determinant has an ambiguity under $\Delta\leftrightarrow\bar{\Delta}$ and so we must include a pole for each. Since the (anti-)quasinormal modes listed above take the same form for $\Delta$ and $\bar{\Delta}$ this effectively doubles the degeneracy of the poles in the one-loop determinant relative to the modes listed in the table.

\subsubsection*{One-loop determinant}

We recall from Sec.\,\ref{section:SpoolDerivation} that for spinning fields we must account for both a set of zero modes \eqref{eq:Zzero}, due to the compact nature of $S^3$, as well as the functional determinant of a symmetric, transverse, traceless field \eqref{eq:ZDs1}. The DHS method can only reproduce this second part of the calculation. Repeating the procedure in the scalar case above for the spin-1 field, the relevant functional determinant for a  transverse vector field is 
\beq
    Z_{\text{STT}} = \prod_{\substack{\pm, n\in\mathbb{N} \\ l,k \in \mathbb{Z}}} 
    \left(
        |k|+2n + |l| + \Delta_{\pm} + 1
    \right)^{-1}
    \prod_{\substack{\pm, \{l>0, k<0\} \\ \cup \{l<0, k>0\}}}
    \left(
        |k|+ |l| + \Delta_{\pm} -1
    \right)^{-1}
\eeq
and hence
\begin{equation}
    \begin{aligned}
         \log(Z_{\text{STT}}) &= -\sum_{\substack{\pm, n\in\mathbb{N} \\ l,k \in \mathbb{Z}}} 
    \log\left(
        |k|+2n + |l| + \Delta_{\pm} + 1
    \right)
   -\sum_{\substack{\pm, \{l>0, k<0\} \\ \cup \{l<0, k>0\}}}
    \log\left(
        |k|+ |l| + \Delta_{\pm} -1
    \right) \\
    &= -\sum_{\pm,p=0}^{\infty} \log(\Delta_{\pm} + 1 + p)(p+1)^2 
    -\sum_{\pm,p=1}^{\infty}\log(\Delta_\pm + p)(2p) \\
    &= -\sum_{\pm,p = 0}^{\infty} \log(\Delta_{\pm} + p)(p+1)^2 
    + \sum_{\pm,p = 0}^{\infty} \log(\Delta_{\pm} + p)~.
    \end{aligned}
\end{equation}
We now identify the first of these sums as exactly $2\log(Z_{\text{scalar}})$ in \eqref{eq:Zs-app}, and for the second term it is convenient to define
\begin{equation}
    \begin{aligned}\label{eq:Zedge-defn}
    \log(Z_{\text{edge}}) &\equiv -\sum_{\pm,p = 0}^{\infty} \log(\Delta_{\pm} + p) \\
    &= \sum_{\pm} \zeta'(0, \Delta_\pm)~.
 \end{aligned}
\end{equation}
We refer to this part of the partition function by the term 'edge' in reference to \cite{Anninos:2020hfj}, where this term is viewed as an 'edge character integral'. Similarly to before the expansion of the zeta function implies that
\beq
\zeta'(0, \pm i \mu)=\zeta'(0, 1 \pm i \mu) - \log(\pm i \mu)~,
\eeq 
and hence
\begin{align}
\begin{split}
    \log(Z_{\text{edge}}) &= \frac{1}{2} \sum_{\pm} \zeta'(0, 1 \pm i \mu)
    +\zeta'(0, \pm i \mu)+\log(\pm i \mu)\\
    &=\frac{1}{2}\sum_{\pm} -\pi i B_1(\pm i\mu) + \text{Li}_1(e^{\pm 2\pi \mu})
    + \log(\pm i \mu)~.
\end{split}
\end{align}
Using the definition of the Bernoulli polynomial and $\Li{1}$ and inserting this expression for $\log(Z_{\text{edge}})$ into the main calculation, we arrive at 
\beq
    \log(Z_{\text{STT}}) = 2\log(Z_{\text{scalar}}) - \left(
        -\pi\mu -\log(1-e^{-2\pi\mu})
    \right) - \frac{1}{2}\log(\mu^2)~.
\eeq
Finally, we need to account for the zero modes. Following \cite{Anninos:2020hfj}, the contribution is given by \eqref{eq:Zzero}, and with this the full one-loop determinant is
\beq
    Z_{\Delta,\mss=1} = Z_{\text{STT}} Z_{\text{zero}} = Z_{\text{STT}} \prod_\pm (\Delta_\pm - 1)^{\frac{1}{2}}~,
\eeq
and so 
\beq
    \log(Z_{\Delta,\mss=1}) = 2\log(Z_{\text{scalar}}) - \left(
        -\pi\mu -\log(1-e^{-2\pi\mu})
    \right)~.
\eeq

\subsection{Massive \texorpdfstring{spin-$\mss$}{spin-s} fields}

It is now clear how the story should continue for massive spin-$\mss$ fields, when $\mss\geq2$. We solve the equations of motion for a spin-$\mss$ field
\beq 
    \nabla^2\Phi_{\mu_1 ... \mu_\mss} = \lds^2\bar{m}_\mss^2 \Phi_{\mu_1 ... \mu_\mss}~,
\eeq
where $\Phi$ is symmetric and subject to the usual constraints
\beq
    \nabla^{\mu_1}\Phi_{\mu_1 ... \mu_\mss}=0,\qquad \Phi^{\nu}_{\ \nu \mu_3 ... \mu_\mss}=0~.
\eeq
Applying (anti-)quasinormal mode boundary conditions we can extract the (anti-)quasinormal mode frequencies.
A careful treatment of these quasinormal modes will show that (as expected) the quasinormal modes which Wick-rotate to regular Euclidean modes on $S^3$ are the symmetric, transverse, traceless spherical harmonics. These lie in representations of $\mathfrak{so}(4)\cong \su\times\su$ labelled by highest weights $(\frac{n}{2}+\mss, \frac{n}{2})$ and $(\frac{n}{2}, \frac{n}{2}+\mss)$.

The corresponding degeneracies for $\Delta=-p, p\geq -\mss$ are $(\mss+1+p)(1-\mss+p)$ and so 
\begin{align}
\begin{split}
    \log(Z_{\text{STT}})
    =& -\sum_{\pm,p=\mss}^{\infty} \log(\Delta_{\pm} + p)(p+1 + \mss)(p+1-\mss) \\
    =& -\sum_{\pm,p=0}^{\infty} \log(\Delta_{\pm} + p)(p+1)^2 + 
    \mss^2\sum_{\pm,p=0}^{\infty} \log(\Delta_{\pm} + p)\\
    &+\sum_{\pm,p=0}^{\mss-2} \log(\Delta_{\pm} + p)(p+1 + \mss)(p+1-\mss)~.
\end{split}
\end{align}
The first of these sums reproduces again $2\log(Z_{\text{scalar}})$ in \eqref{eq:logZs-app} and the second sum is proportional to $\log(Z_{\text{edge}})$ as defined in \eqref{eq:Zedge-defn}. We can directly insert our previous expression for this.
\begin{align}
\begin{split}
    \log(Z_{\text{STT}})=& 2\log(Z_{\text{scalar}}) - \mss^2\left(
        -\pi\mu -\log(1-e^{-2\pi\mu})
    \right) - \frac{\mss^2}{2}\log(\mu^2)\\
     &+\sum_{\pm,p=0}^{\mss-2} \log(\Delta_{\pm} + p)(p+1 + \mss)(p+1-\mss)~.
\end{split}
\end{align}
As in the spin-1 case, to compute the full local partition function we also need to account for a set of zero modes \eqref{eq:Zzero}. Including these, we obtain the final result
\beq
    \log(Z_{\Delta,\mss})=\log(Z_{\text{STT}})+\log(Z_{\text{zero}}) = 
    2\log(Z_{\text{scalar}}) - \mss^2\left(
        -\pi\mu -\log(1-e^{-2\pi\mu})
    \right)~.
\eeq
This formula is again in agreement with the calculation in \cite{Anninos:2020hfj}, and it matches with the Wilson spool in \eqref{eq:TreeResult} (up to overall phase).

\bibliographystyle{ytphys}
\bibliography{dSWilsonRefs.bib}

\end{document}